\documentclass[twocolumn,aps,prb,showpacs,floatfix,epsfig]{revtex4}

\usepackage{graphicx}%
\usepackage{dcolumn}
\usepackage{bm}%

\newcommand \be{\begin{equation}}
\newcommand \ee{\end{equation}}
\newcommand \ba{\begin{eqnarray}}
\newcommand \ea{\end{eqnarray}}
\begin{document}
\title
{\bf Theory of superfast fronts of impact ionization in semiconductor structures}

\author{Pavel Rodin \cite{EMAIL}}
\affiliation
{Ioffe Physicotechnical Institute of Russian Academy of Sciences, Politechnicheskaya 26,
194021, St.-Petersburg, Russia}

\author{Ute Ebert}
\affiliation
{Centrum voor Wiskunde en Informatica, Postbus 94079, 1090 GB Amsterdam,
The Netherlands}

\author{Andrey Minarsky}
\affiliation
{Physico-Technical High School of Russian Academy of Sciences,
Khlopina 8-3,194021, St.-Petersburg, Russia}

\author{Igor Grekhov}
\affiliation
{Ioffe Physicotechnical Institute,
Politechnicheskaya 26, 194021, St.-Petersburg, Russia}

\setcounter{page}{1}
\date{\today}

\hyphenation{cha-rac-te-ris-tics}
\hyphenation{se-mi-con-duc-tor}
\hyphenation{fluc-tua-tion}
\hyphenation{fi-la-men-ta-tion}
\hyphenation{self--con-sis-tent}
\hyphenation{cor-res-pon-ding}
\hyphenation{con-duc-ti-vi-ti-tes}


\begin{abstract}
We present an
analytical theory for impact ionization fronts in reversely biased
$p^{+}$-$n$-$n^{+}$ structures. The front propagates into a depleted $n$
base  with a velocity that exceeds the saturated drift velocity. The
front passage generates a dense electron-hole plasma and in this way switches
the structure from low to high conductivity. For a
planar front we determine the concentration of the generated plasma,
the maximum electric field, the front width and the voltage over the $n$
base as functions of front velocity and doping of the $n$
base. Theory takes into account that drift velocities and impact ionization
coefficients differ between  electrons and
holes, and it makes quantitative predictions for any semiconductor
material possible.
\end{abstract}



\maketitle

\section{Introduction}

Fronts of impact ionization can be excited in layered
semiconductor  structures such as $p^{+}$-$n$-$n^{+}$ diodes and
$p^{+}$-$n$-$p$-$n^{+}$ dynistors.
\cite{PRA68,DEL70,GRE79,GRE81,GRE81a,BEN85,ALF87,EFA88,EFA90} The
front passage fills the structure with dense electron-hole plasma
and hence leads to the transition of the  reversely biased $p$-$n$
junction from low-conducting to high conducting state. Apart from
microwave TRAPATT (TRApped Plasma Avalanche Triggered Transit)
diodes \ \cite{PRA68,DEL70}  triggering of impact ionization
fronts has been observed in high voltage $p^{+}$-$n$-$n^{+}$
diodes manufactured from both Si \ \cite{GRE79,GRE81,GRE81a,BEN85}
and GaAs. \cite{ALF87,EFA88,EFA90} In modern semiconductor
electronics excitation of ionization front  is a unique nonoptical
method capable to form subnanosecond voltage ramps with kilovolt
amplitudes. It has found numerous pulse power applications.
\cite{applications1, applications2}

The mechanism of front propagation is based on avalanche
multiplication of carriers by impact ionization and subsequent
screening of the ionizing electric field due to the Maxwellian
relaxation in the  generated electron-hole plasma. The qualitative
picture of the front passage is well-known (e.g., see Ref.\
\onlinecite{LEV05} and references therein). In Fig.\ \ref{sketch1}
we sketch the profiles of the electric field $E$ and the total carrier
concentration $\sigma=n+p$ (here $n$ and $p$ are concentrations of
electron and hole, respectively) in the $n$ base of reversely biased
$p^{+}$-$n$-$n^{+}$ diode structure. The front propagates into
a {\it depleted region} where the concentration of free carriers
$\sigma_0$ is much smaller than the concentration of dopants
$N_d$. In the depleted region the slope of electric field $qN_d/
\varepsilon \varepsilon_0$ is controlled by the charge of ionized
donors. The {\it ionization zone} travels into the depleted region
with velocity $v_f$ that exceeds the saturated drift velocity $v_{ns}$ of
electrons. This is possible due to a small concentration of
free carriers $\sigma_0$ in the depleted region. The 
multiplication of these carriers starts as soon as electric field becomes
sufficiently strong, therefore the ionization zone can propagate
faster than the drift velocity. Generation of electron-hole pairs and subsequent separation of
electron and hole in the electric field form a {\it screening region} behind the
ionization zone. Here the drift velocities remain saturated. The
space charge in the screening region is due to excessive electron
concentration. In the {\it plasma} layer the carrier concentration
$\sigma_{\rm pl}$ exceeds $N_d$ by several orders of magnitude,
electric field is low and corresponds to linear Ohmic regime.
The described propagation of ionization front resembles the propagation
of finger-like streamers into pre-ionized medium 
(see discussion in Ref.\ \onlinecite{ROD02}).  

The analytical theory of traveling ionization fronts in
$p^{+}$-$n$-$n^{+}$ structures has been pioneered in Ref.\
\onlinecite{DEL70} for TRAPATT diodes.  This classical work is
based on two crucial simplifications that make the theory
essentially qualitative: (i) impact ionization coefficients are
modeled by step functions and (ii) electrons and holes are assumed
to be identical. Later on the focus mostly shifted from analytical
studies to numerical simulations.  \cite{BIL83,KAR96,GAU98}
The demand for a quantitative analytical description remains strong,
in particular, as ionization fronts in wide band materials \ \cite{ROD06} 
and also operation at electric fields above the band-to-band Zener
breakdown\ \cite{ROD02a,RUK05} have a promising prospective.
Analytical theory  is also important
to place ionization fronts in doped semiconductor structures into
the general context of studies on front dynamics in spatially
extended nonlinear systems. \cite{MEE78,PEL88,CRO93}

This article presents a theory
of ionization front in a reverserly biased  $p^{+}$-$n$-$n^{+}$ structure.
We take the asymmetry
between electron and holes in both transport and impact ionization into account and assume
a general form of impact ionization coefficients.
The theory determines the maximum electric field in the traveling front $E_{\rm m}$,
the voltage over the structure $u$,
the concentration of the generated plasma $\sigma_{\rm pl}$
and the electric filed in plasma $E_{\rm pl}$ as functions
of front velocity $v_f$ and the $n$ base doping $N_d$ for self-similar propagation
with constant velocity $v_f$.
These results determine the instant front velocity $v_f$ as
a function of the applied voltage $u$ and front position $x_f$ in cases when
the front velocity and shape vary during the passage.

\section{The model}

\subsection{Basic equations in drift-diffusion approximation}

We investigate how an impact ionization front passes through a uniformly dope $n$ base of 
a reversely biased  $p^{+}$-$n$-$n^{+}$ diode
structure  as sketched in Fig.\ \ref{sketch1}.
Heavily doped $p^{+}$ and $n^{+}$ layers play the role of contacts
and are not taken further into consideration. The carrier dynamics
in the $n$ base is described by the standard set of  continuity
equations and the Poisson equation
\begin{eqnarray}
\label{continuity1}
&&\partial_t \, n - \partial_x \left[ v_n(E)\cdot n \right] - D_n \, \partial_x^2 \, n=G(n,p,E),\\
\nonumber
&&\partial_t \, p + \partial_x \left[ v_p(E)\cdot p \right] - D_p \, \partial_x^2 \, p=G(n,p,E),\\
\nonumber
&&\partial_x E = \frac{q}{\varepsilon \varepsilon_0} (p-n+N_d),
\end{eqnarray}
where $n,p$ and $v_n,v_p$ are electron and holes concentrations and  drift velocities,
respectively; $E$ is the electric  field strength, $N_d={\rm const}$ is concentration of
donors in the $n$ base, $q > 0$ is the elementary charge,
$\varepsilon$ and $\varepsilon_0$ are the permittivity of the material and the absolute
permittivity, respectively.  We use the notation
$v_{n,p}(E)>0$ and take the actual direction of the carrier drift
through the signs in equations (\ref{continuity1}) into account. The impact ionization term
is given by
\begin{equation}
\label{impactTerm1}
G(n,p,E)= \alpha_n(E) \, v_n(E) \, n + \alpha_p (E) \, v_p(E) \, p,
\end{equation}
where $\alpha_{n,p}(E)$ are impact ionization coefficients.

Let us introduce new variables
\begin{equation}
\label{variables}
\sigma \equiv n+p, \qquad \rho \equiv p-n.
\end{equation}
The first variable $\sigma$ is the total concentration of free carriers,
the second one $\rho$ is proportional to the space charge of
free carriers. Neglecting diffusion that does not play any role on the relevant
scales (e.g., see Ref.\ \onlinecite{ROD02}), we present Eqs.\ (\ref{continuity1}) as
\begin{eqnarray}
\label{sigma1}
&&\partial_t \, \sigma + \partial_x \left[v^{-}(E)\,\sigma + v^{+}(E) \, \rho \right]=2\,G,
\\ \label{rho1}
&&\partial_t \, \rho + \partial_x \left[v^{+}(E)\,\sigma + v^{-}(E) \, \rho \right]=0,
\\ \label{field1}
&& \partial_x E= \frac{q}{\varepsilon \varepsilon_0}\left[\rho + N_d \right],
\end{eqnarray}
where
\begin{equation}
v^{\pm}(E) \equiv \frac{v_p(E) \pm v_n(E)}{2}.
\end{equation}
Note that for most semiconductors $v^{-}(E)<0$.

Solving Eq.\ (\ref{field1}) for $\rho$ , substituting this $\rho$ into the expression $\partial_t \rho$ 
in Eq.\ (\ref{rho1}) and integrating
over $x$, we obtain the conservation of the total current density in a one-dimensional system
\begin{equation}
\label{current1}
J = q \left[v^{+}(E)\, \sigma + v^{-}(E)\, \rho \right]+\varepsilon \varepsilon_0 \, \partial_t E \qquad \partial_x J=0.
\end{equation}
Here the first and the second term correspond to the conduction and displacement components
of the current density $J$.
In the following we replace Eq.\ (\ref{rho1}) by Eq.\ (\ref{current1}).

\subsection{Self-similar propagation of the ionization front}

We consider the fronts moving in the same direction as electrons drift
- so called {\it negative} fronts.
We choose $E>0$, hence electrons and negative fronts move to the left, cf.\ Fig.\ \ref{sketch1}.
{\it Positive} fronts in $n^{+}$-$p$-$p^{+}$ structures that move in the same
direction as hole drift, can be described by exchanging electrons
and holes in Eqs.\ (\ref{continuity1}) and replacing the donor concentration $N_d$ by the same
acceptor concentration $N_a$. For the self-similar front motion with velocity
$v_{f} = {\rm const}$ we get
\begin{eqnarray}
\nonumber
\sigma(x,t)&=&\sigma(x+v_f \, t),\\
\nonumber
\rho(x,t)&=&\rho(x+v_f \, t), \\
\nonumber
E(x,t)&=&E(x+v_f \, t), 
\end{eqnarray}
where we fixed the notation such that the fronts move with a positive velocity $v_f>0$ in the negative
$x$ direction.
Then in the comoving frame $z \equiv x + v_f \, t$ Eqs.\ (\ref{sigma1}),(\ref{current1}) and (\ref{field1})
become
\begin{eqnarray}
\label{sigma2}
&&d_z \left[(v_{f}+v^{-}) \sigma+v^{+} \rho \right]= 2\, G,
\\ \label{current2}
&&J = q \left[v^{+}\, \sigma+v^{-}\, \rho \right]+\varepsilon \, \varepsilon_0 \,v_{f} \, 
d_z E, \; d_z J=0
\\ \label{field2}
&& d_z E= \frac{q}{\varepsilon \varepsilon_0} \left[\rho + N_d \right].
\end{eqnarray}
The velocity of the negative front  $v_f$ can not be smaller
than the saturated electron velocity $v_{ns}$. \cite{DEL70,LEV05}
As we will see in the next section, a constant velocity $v_f = {\rm const}$ implies a time independent
total current density $J = {\rm const}$.

\subsection{Relation between the current density and the front velocity}

Using Eq.\ (\ref{field2}) to eliminate $d_z E$ from Eq.\ (\ref{current2}), we find
\begin{equation}
\label{current3}
d_z J=0, \; J = q\, v_f \, N_d + q \,j,
\; j \equiv v^{+} \sigma + \left[v_{f}+v^{-}\right]\rho.
\end{equation}
Here $j$ is chosen to be a particle current density whereas $J$ is a charge current density.
Both $J$ and $j$ are constants in space. Typically
the electric field at the left boundary $E_{\rm left} \equiv E(x=0)$ is too low for impact ionization
[$\alpha_{n,p}(E_{\rm left})=0$]  but sufficiently strong to saturate the drift velocities so that
$v_n(E_{\rm left})=v_{ns}$, $v_n(E_{\rm left})=v_{ps}$ (see Fig.\ \ref{sketch1}).
Then $j$ is determined by small
concentrations $n_0,p_0 \ll N_d$ of free carriers that are present during the front passage
in the depleted part of the structure far away from the ionization zone:
\begin{eqnarray}
\label{leak}
&&j = v^{+}_{s} \, \sigma_0 + \left[v_{f}+v^{-}_{s}\right]\rho_0, \\
\nonumber
&&\sigma_0 \equiv n_0+p_0, \; \rho_0 \equiv p_0-n_0, \;
v^{\pm}_{s}  \equiv \frac{v_{ps} \pm v_{ns}}{2}.
\end{eqnarray}
Since $\sigma_0,\rho_0 \ll N_d$ the second term in Eq.\ (\ref{current3})
is negligible.
Hence $J \approx q N_d \, v_f$. This result is well known \ \cite{DEL70}
and physically clear: since the front propagates with
velocity $v_f$ into a charged medium with space charge density $q N_d$,
the current density $q N_d \, v_f$
is required to neutralize the space charge of the ionized donors.
Hence for a planar front, self-similar propagation with constant velocity
$v_f$ corresponds to a fixed total current in the external circuit
$d_t J=0$.

Although the initial carrier density in the depleted region is low $\sigma_0 \ll N_d$,
their presence is a necessary requirement for the front to propagate with a velocity $v_f$ that exceeds the electron drift velocity $v_{ns}$. This fast propagation mode is possible because these initial carriers mulitply in an avalanche like manner as soon as the electric field exceeds the ionization threshold.

In the case under study the concentrations $n_0$ and
$p_0$ are not permanently nonvanishing in the medium  the front
propagates into. On the contrary, a certain mechanism creates these
carriers in the $n$ base just before the front starts to travel,
and it actually triggers this propagation. This mechanism is not
universal and depends on the specific design and operation mode of a
semiconductor device. In microwave TRAPPAT diodes the carriers
that remain in the structure from the previous front passage serve
as initial carriers for the next passage. \cite{LEV05} In
contrast, in high voltage diodes used as pulse sharpeners the time
period between subsequent front passages is so long (typically $ >
100 \; {\rm \mu s}$) that each front passage represents
an independent event. \cite{GRE79,GRE81,GRE81a} Between the pulses
the reverse voltage is kept close to the stationary breakdown
voltage $u_b$. During this waiting period that can be arbitrarily
long the leakage current is much smaller than $q\,j$, so that $n$
base is essentially empty. To trigger the front the applied
voltage $u$ is being rapidly increased above $u_b$. Experiments
show that the front starts to travel when $u$ exceeds $u_b$
several times. \cite{GRE79,GRE81,GRE81a} It appears that in these
devices the initial carriers are generated by field-enhanced
ionization of deep-level electron traps. \cite{AST98,APL05,JAP05}
The release of electrons bound on these deep centers triggers the
front and provides conditions for its superfast propagation.
\cite{footnote1} Here we focus exclusively on the stage when
the ionization front is already traveling and refer to Refs.\
\onlinecite{APL05,JAP05} for the detailed discussion on triggering
and initial carriers problems. We assume that concentration of
initial carriers suffices for using our density model and
we treat $\sigma_0$ as an input parameter of our model.

\subsection{The final set of equations}

Eq.\ (\ref{current3}) allows to express the ``space charge'' $\rho$ via the
total concentration $\sigma$ as
\begin{equation}
\label{rho2}
\rho= - \frac{v^{+}(E)}{v_f+v^{-}(E)} \sigma + \frac{j}{v_f+v^{-}(E)} \qquad d_z j=0.
\end{equation}
Substituting Eq.\ (\ref{rho2}) in Eq.\ (\ref{sigma2}), we obtain the final differential equation
for $\sigma$
\begin{equation}
\label{sigma3}
d_z \left[\frac{(v_{f}+v^{-})^2-(v^{+})^2}{v_f+v^{-}} \, \sigma+\frac{v^{+}}{v_f+v^{-}} \, j \right]=2G.
\end{equation}
The impact ionization can be expressed via variables
$\sigma$ and $\rho$ as
\begin{equation}
\label{impactTerm2}
G(\sigma,\rho,E)= \beta^{+}(E) \, \sigma + \beta^{-}(E) \, \rho
\end{equation}
with impact ionization frequencies
\begin{equation}
\label{frequencies}
\beta^{\pm} \equiv \frac{v_p(E) \, \alpha_p (E) \pm v_n(E) \, \alpha_n(E)}{2}
\end{equation}
Using Eq.\ (\ref{rho2}) again to exclude $\rho$, we obtain
\begin{eqnarray}
G(\sigma,E) &=& \beta_{\rm eff}(E,v_f) \, \sigma+\frac{\beta^{-}}{v_f+v^{-}}\,j, \\ \nonumber
&&\beta_{\rm eff}(E,v_f) \equiv \frac{\beta^{+}\left[v_f+v^{-}\right]-\beta^{-}v^{+}}{v_f+v^{-}} \, .
\end{eqnarray}
Finally, we express $\rho$ by $\sigma$ in the Poisson equation (\ref{field2}).
This leads to Eq.\ (\ref{field3}) below.

The two equations for the total carrier concentration $\sigma$ and the field $E$
\begin{eqnarray}
\label{sigma4}
d_z \left[\frac{(v_{f}+v^{-})^2-(v^{+})^2}{v_f+v^{-}} \, \sigma + \frac{v^{+}}{v_f+v^{-}} \, j \right]&=&
\\ \nonumber
= 2 \, \beta_{\rm eff}(E,v_f) \, \sigma &+& \frac{2 \beta^{-}}{v_f+v^{-}}\,j \,,
\end{eqnarray}
\begin{eqnarray}
\label{field3}
d_z E = b - \frac{q}{\varepsilon \varepsilon_0}  \frac{v^{+}}{v_f+v^{-}} \, \sigma&,& \\
\nonumber
&b& \equiv \frac{q}{\varepsilon \varepsilon_0} \, \left( N_d + \frac{j}{v_f+v^{-}}\right), 
\end{eqnarray}
with the known functions $v^{\pm}(E)$, $\beta^{\pm}(E)$ and $\beta_{\rm eff}(E,v_f)$ and the constant
\begin{equation}
\label{current4}
j \equiv v^{+} \sigma + \left[v_{f}+v^{-}\right]\rho.
\end{equation}
completely describes the self-similar
propagation of impact ionization front.

Summarizing, we have first substituted the variables $(n,p,E)$ by the new variables $(\sigma,\rho,E)$,
and then we have expressed $\rho$ by the conserved total current $J$(or $j$) and $\sigma$. In this way we have eliminated the third equation front the original set
(\ref{continuity1}) using the fact that the local space charge density $\rho$ is no independent variable when the conserved total current $J$ and the local carrier density $\sigma$ and field $E$ are fixed.
The form of the final equations (\ref{sigma4}) and (\ref{field3}) is one of two odinary differential equations of first order.

\subsection{Special and limiting cases}

Apart from the general case $v_n(E) \ne v_p(E)$, $\alpha_n(E)  \ne \alpha_p(E)$,
it is instructive to consider the following special cases:
\begin{eqnarray}
\nonumber
({\rm a}) \;  v_n(E)=v_p(E) \equiv v(E), &&  \alpha_n(E)=\alpha_p(E) \equiv \alpha(E) ,\\
\nonumber
({\rm b}) \;  v_n(E)=v_p(E) \equiv v(E), &&  \alpha_p(E)=0,  \;  \alpha_n(E) \equiv \alpha(E),\\
\nonumber
({\rm c}) \;  v_n(E)=v_p(E) \equiv v(E), &&  \alpha_n(E)=0, \; \alpha_p(E) \equiv \alpha(E).
\end{eqnarray}
In case (a) electrons and holes are fully equal, in cases (b,c) their drift velocities are equal,
but impact ionization takes place due to only one type of carriers. For example, cases (b) and (c)
are simplified but reasonable approximations for Si and SiC, \ \cite{handbook} respectively.

Another two limiting cases provide insight in the front dynamics for materials where the 
asymmetry between positive and negative carriers is very strong:
\begin{eqnarray}
\nonumber
({\rm d}) \;  v_n(E) \equiv v(E), \; v_p(E)=0,\;&&  \alpha_n(E) \equiv \alpha(E), \;
\alpha_p=0,\\
\nonumber
({\rm e}) \;  v_p(E) \equiv v(E), \; v_n(E)=0, \;&&  \alpha_p(E) \equiv \alpha(E),
\; \alpha_n = 0.
\end{eqnarray}
Below we refer to these special cases as to cases (a),(b),(c),(d) and (e), also
to denote the respective curves in figures.

\section{General properties of the stationary front propagation}

The main parameters  of the traveling front are the maximum electric
field $E_{\rm m}$, the width $\ell_{\rho}$ of the screening region,
the voltage $u$ across the $n$ base, the carrier
concentration $\sigma_{\rm pl}$ and electric field $E_{\rm
pl}$ in the electron-hole plasma behind the front (Fig.\
\ref{sketch1}). [In Fig.\ \ref{sketch2} we sketch the respective $\sigma(E)$
dependence that follows from
$E(x)$ and $\sigma(x)$ dependences shown in Fig.\ \ref{sketch1}.]
In this section we relate these parameters to
the front velocity $v_f$, the doping level $N_d$ and the initial
concentration $\sigma_0$ in the deleted region, assuming that
$v_f = {\rm const}$ and front propagation is self-similar. Note that
according to Eq.\ (\ref{current3}), the velocity $v_f$ can be
expressed by the current density $J=q N_d \, v_f$.

\subsection{Ordering of scales}

In semiconductors the drift velocities $v_{n,p}(E)$ typically
saturate in electric fields  above the characteristic fields
$E_{ns,ps}$ that are much smaller than the effective threshold of
impact ionization. Consequently, in the range of electric fields
where impact ionization sets in ($\alpha_{n,p} \ne 0$) the
velocities of free carries do not depend on electric field
[$v^{+}(E)=v^{+}_{s}$, $v^{-}(E)=v^{-}_{s}$ in Eqs.\
(\ref{sigma4},\ref{field3},\ref{current4})]. On the other hand,
the generation term is negligible in the range of electric fields
where the nonlinearity of $v_{n,p}(E)$ dependencies is essential.

For semiconductors where the drift velocities
$v_{n}(E)$ and $v_{p}(E)$ depend monotonically on the field, the approximations
\begin{equation}
\label{velocityDrift}
v_{n,p}(E)=v_{ns,ps}\frac{E}{E+E_{ns,ps}}
\end{equation}
and their modifications are widely used. \cite{JAC77}
The impact ionization coefficients $\alpha_n(E)$ and $\alpha_p(E)$
for electrons and holes are usually modeled as\ \cite{Sze}
\begin{equation}
\label{ionizationCoeff}
\alpha_{n,p}(E)=\alpha_{n0,p0}\exp \left(-\frac{E_{n0,p0}}{E}\right).
\end{equation}
This approximation is known as Townsend approximation in gas discharge
physics. \cite{Raizer} The characteristic transport fields $E_{ns,ps}$
are typically much smaller than the impact ionization fields $E_{n0,p0}$:
$E_{ns,ps} \ll E_{n0,p0}$.
For example, in Si, we have $E_{ns,ps} \sim 10^4$~V/cm whereas $E_{n0,p0} \sim 10^6$~V/cm.
\cite{handbook} It should be noted that none of these explicit approximations
(\ref{velocityDrift}) or (\ref{ionizationCoeff}) are needed for our
analytical results.

Eqs.\ (\ref{sigma4}),(\ref{field3}) and (\ref{current4}) will be solved separately
in the range of strong electric fields $E_s < E \lesssim E_0$ where impact ionization
takes place and the range of moderate-to-low electric field $E\lesssim E_s$
where transition from high-field
transport to low-field transport occurs and the plasma layer is formed.
Due to the overlap between these two regions the obtained solutions
can be sewn together providing a consistent description.

\subsection{Equations in the high field region}

For $E>E_{ns,ps}$ we assume that carriers drift with constant velocities
$v_{n,p}(E)=v_{ns,ps}$. Then Eqs.\ (\ref{sigma4},\ref{field3},\ref{current4})
become
\begin{eqnarray}
\label{sigmaHigh}
&&d_z \sigma = \lambda \, \beta_{\rm eff}(E,v_f) \, \sigma,
\\ \label{fieldHigh}
&&d_z E = b -c \, \sigma,
\\ \label{currentHigh}
&&j=v^{+}_{s} \, \sigma+(v_f+v^{-}_{s})\rho,
\end{eqnarray}
where
\begin{eqnarray}
\label{coefficientsHigh}
&&\lambda \equiv \frac{2(v_f+v^{-}_{s})}{(v_f+v^{-}_{s})^2-(v^{+}_{s})^2}, \\
&&b \equiv \frac{q}{\varepsilon \varepsilon_0} \left(N_d+\frac{j}{v_f+v^{-}_{s}}\right)\approx
\frac {q N_d}{\varepsilon \varepsilon_0}, \\ \nonumber
&&c \equiv \frac {q}{\varepsilon \varepsilon_0} \frac{v^{+}_{s}}{v_{f}+v^{-}_{s}}, \qquad
v^{\pm}_{s} \equiv \frac{v_{ps} \pm v_{ns}}{2}.
\end{eqnarray}
The coefficients $\lambda$, $b$ and $c$ are constants.
When deriving Eq.\ (\ref{sigmaHigh}) from Eq.\ (\ref{sigma4}) we neglect the second term
on the right-hand side. This is justified because
the order of magnitude value of this term is
$\beta^{-} \, (v^{+}_{s}/v_f)\sigma_0$ which is much smaller than the first term.
We also take into account that $j/(v_f+v^{-}) \sim (v_s/v_f)\sigma_0 \ll N_d$
in the expression for $b$.

\subsection{Carrier concentration just behind the ionization zone}

At the point $x_f$ where electric field reaches the maximum value
$E_{\rm m}$ the slope of electric field profile
is equal to zero (see Fig.\ \ref{sketch1}). In the comoving frame $x_f$ corresponds
to the point $z_{\rm m}=x_f+v_f \, t$ that does not vary in time.
It follows from $d_z E=0$ and Eq.\ (\ref{fieldHigh}) that
\begin{eqnarray}
\label{sigma_m}
\sigma_{\rm m} \equiv \sigma(z_{\rm m})
= \frac{b}{c}&=& \\ \nonumber
&=& \frac{v_f+v^{-}_{s}}{v^{+}_{s}}\,N_d+\frac{j}{v^{+}_{s}}
\approx \frac{v_f+v^{-}_{s}}{v^{+}_{s}} \, N_d.
\end{eqnarray}
Dividing Eq.\ (\ref{sigmaHigh}) by Eq.\ (\ref{fieldHigh}) we exclude $z$,
and then employ (\ref{sigma_m}). This yields
\begin{equation}
\label{field_sigma}
\frac{\sigma_{\rm m} - \sigma}{\sigma} \, d \sigma=
\frac{\lambda}{c} \, \beta_{\rm eff}(E,v_f) \, dE.
\end{equation}
Let us denote as $\sigma^{\star}$ the concentration $\sigma$ which is reached
just behind the ionization zone (Fig.\ \ref{sketch1}).
Then the integrals of the left-hand side of  Eq.\ (\ref{field_sigma})
from $\sigma_0$ to $\sigma_{\rm m}$
and from $\sigma^{\star}$ to $\sigma_{\rm m}$ are both equal to the integral
of the right-hand side from 0 to $E_{\rm m}$ (see Fig.\ \ref{sketch2}):
\begin{eqnarray}
\label{FirstIntegral1}
\int_{\sigma_0}^{\sigma_{\rm m}} \frac{\sigma_{\rm m} - \sigma}{\sigma} d \sigma &=&
\int_{\sigma_{\star}}^{\sigma_{\rm m}} \frac{\sigma_{\rm m} - \sigma}{\sigma} d \sigma=\\
\nonumber
&=&\frac{\lambda}{c} \int_0^{E_{\rm m}} \, \beta_{\rm eff}(E,v_f)  \, dE.
\end{eqnarray}
Taking integrals over $\sigma$, we find the relation
between the concentration behind the front $\sigma^{\star}$, front velocity
$v_f/v_s$, the doping level $N_d$ and initial concentration $\sigma_0$:
\begin{equation}
\label{sigma_star}
\sigma^{\star}= \sigma_{\rm m} \ln\frac{\sigma^{\star}}{\sigma_0}+\sigma_0 \approx
\sigma_{\rm m} \ln\frac{\sigma^{\star}}{\sigma_0}.
\end{equation}
The concentration $\sigma^{\star}$ does not depend on $\beta_{\rm eff}$ and thus
on the specific form of $\alpha_{n,p}(E)$.
Equation \ (\ref{sigma_m}) generalizes
equation (29) in Ref.\ \onlinecite{DEL70}.

\begin{figure*}[hp]
\begin{center}
\includegraphics[width=8.5 cm,height=12.0 cm,angle=270]{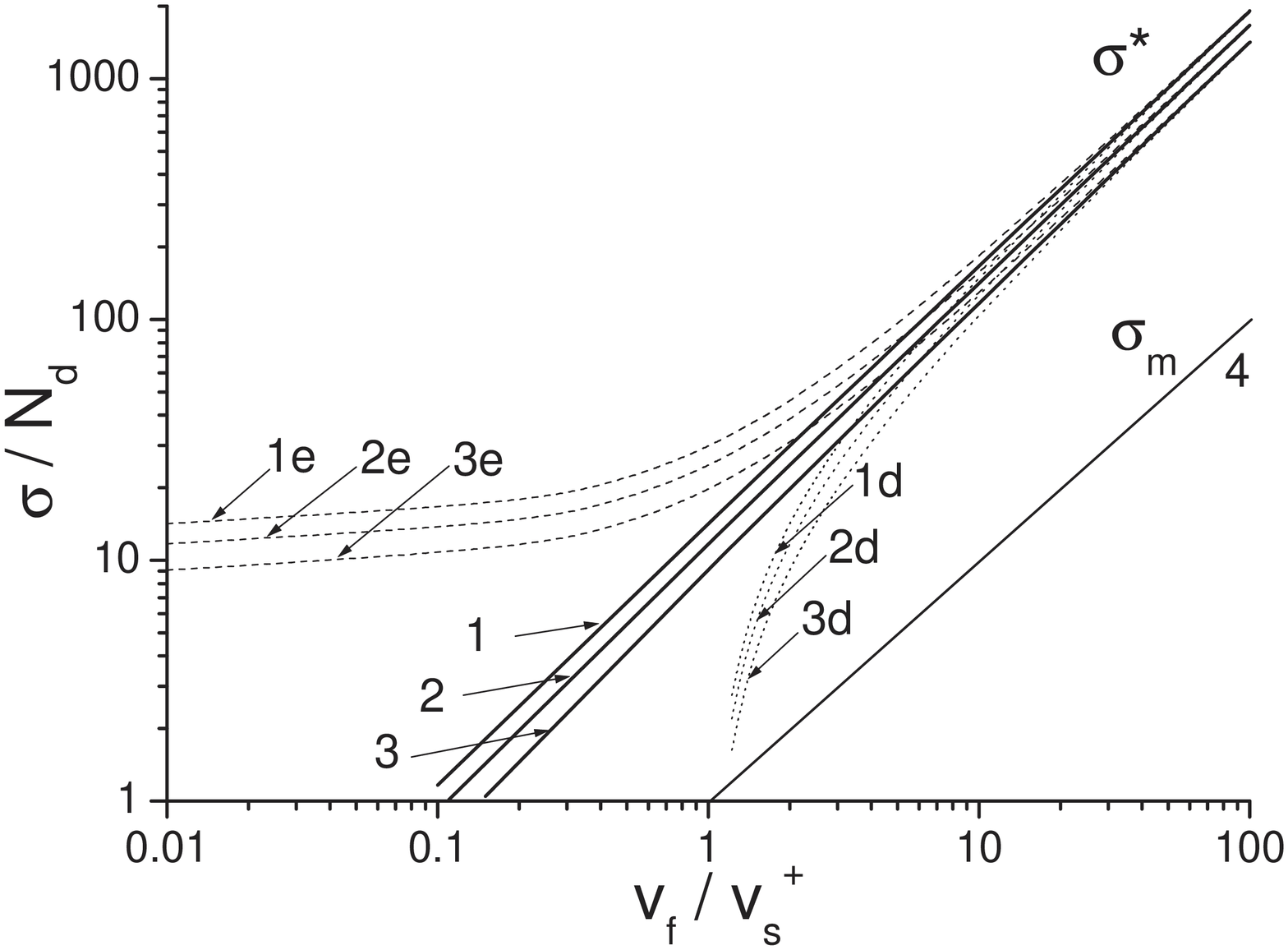}
\end{center}
\caption
{Concentration $\sigma^{\star}$ just behind the ionization zone as a function
of  $v_f/v_s^{+}$ for different values of $\sigma_0/N_d$. 
Thick solid lines 1,2,3 correspond to symmetric case $v^{-}_s=0$.
Dotted lines 1d,2d,3d and dashed lines 1e,2e,3e
correspond to the two limiting asymmetric
cases $v^{-}_{s}/v^{+}_{s} = -1$ [immobile holes, case (d)] and
$v^{-}_{s}/v^{+}_{s} = 1$ [immobile electrons, case(e)], respectively.
Curves of 1st, 2nd and 3rd series correspond to
$\sigma_0/N_d=10^{-3},10^{-4},10^{-5}$, respectively.
Thin solid line 4 shows concentration
$\sigma_{\rm m}$ at the point of maximum electric field
for the symmetric case $v^{-}_s=0$.}
\label{sigmaStar}
\end{figure*}

In Fig.\ \ref{sigmaStar} we show $\sigma^{\star}$ and $\sigma_{\rm
m}$ as functions of $v_f/v^{+}_s$ for different values of
$\sigma_0 / N_d$. Solid lines 1, 2, 3 correspond to the symmetric
case $v^{-}_s=0$. These dependencies are valid for all three
special cases (a,b,c) (see Sec.\ 2E) if $v^{+}_{s}$ is replaced by
$v_s$.  We see that $\sigma^{\star}$ is approximately 10 times
larger than $\sigma_{\rm m}$ (line 4) whereas $\sigma_{\rm m}$
exceeds $\sigma_0$ by several orders of magnitude. Dashed lines
1d, 2d, 3d and dotted lines  1e, 2e, 3e show $\sigma^{\star}$ for
two extreme asymmetric cases $v^{-}_s/v^{+}_s= \mp 1$ that
correspond to immobile holes [case (d), $v_{ps}=0$] and immobile
electrons [case (e), $v_{ns}=0$], respectively. In the case (d) of
immobile holes $\sigma^{\star} \rightarrow 0$  when $v_f/v^{+}_s
\rightarrow 1$ (although the part $v_f < 2 v^{+}_s = v_{ns}$ is
unphysical). In contrast, in the case (e) of immobile electrons
$\sigma^{\star}$ practically does not depend on $v_f$ for
$v_f/v_s^{+}<1$. This plateau exists for $v_s^{-}>0$ and
corresponds to the interval of front velocities $v_{ns}< v_f <
v_{ps}$ (recall that for negative front $v_f \ge v_{ns}$). Even
for the limiting  cases $v^{-}_s/v^{+}_s= \mp 1$ the effect of
transport asymmetry is small for $v_f/v^{+}_s > 10$. Actual values
of $|v^{-}_{s}/v^{+}_{s}|$ are much smaller: at room temperatures
we have $v^{-}_{s}/v^{+}_{s} \approx -0.1, -0.05, -0.3$
for Si, GaAs  and SiC, respectively. \cite{handbook}

According to Eq.\ (\ref{currentHigh}) the space charge of free carriers
behind the ionization zone $\rho^{\star}$ is proportional to $\sigma^{\star}$:
\begin{equation}
\label{rho_star}
\rho^{\star}=-\frac{v^{+}_s}{v_f+v^{-}_s} \, \sigma^{\star}+ \frac{j}{v_f+v^{-}_s}
\approx -\frac{v^{+}_s}{v_f+v^{-}_s} \, \sigma^{\star}
\end{equation}
Electron and holes concentrations
$n^{\star},p^{\star} = (\sigma^{\star} \mp \rho^{\star})/2$ are recovered as
\begin{eqnarray}
\label{np_star}
n^{\star}= \frac{v_f+v_{ps}}{2 \, v_f+v_{ps}-v_{ns}} \, \sigma^{\star}, \\
\nonumber
p^{\star}= \frac{v_f-v_{ns}}{2 \, v_f+v_{ps}-v_{ns}} \, \sigma^{\star}.
\end{eqnarray}
The appearance of negative space charge $\rho^{\star}=p^{\star}-n^{\star}<0$
is a combined effect of spatially inhomogeneous ionization and separation
of electrons and holes in strong electric field.

\subsection{Maximum electric field}

The maximum
field $E_{\rm}$ is determined by integrating Eq. (\ref{field_sigma}):
\begin{equation}
\label{E_m_1}
\int^{E_{\rm m}}_0 \beta_{\rm eff}(E)dE =
\frac{b}{\lambda}\left(\ln \frac{\sigma_{\rm m}}{\sigma_0}-1+\frac{\sigma_0}{\sigma_{\rm m}}\right).
\end{equation}
Substituting the expressions for $\beta_{\rm eff}(E)$, $\lambda$, $b$ and $\sigma_{\rm m}$,
we obtain the explicit formula
\begin{eqnarray}
\label{E_m_2}
&&\frac{v^{+}_s-v^{-}_s}{v_f+v^{-}_{s}-v^{+}_s} \int_0^{E_{\rm m}} \alpha_n(E) \, dE +
\\ \nonumber
&& \qquad \qquad +\frac{v^{+}_s+v^{-}_s}{v_f+v^{-}_{s}+v^{+}_{s}} \int_0^{E_{\rm m}} \alpha_p(E) \, dE =
\\ \nonumber
&& \qquad \qquad \qquad \qquad =\frac{q N_d}{\varepsilon \varepsilon_0}
\left[\ln \left(\frac{v_f+v^{-}_{s}}{v^{+}_{s}}\frac{N_d}{\sigma_0}\right)-1\right].
\end{eqnarray}
Straightforward numerical integration of Eq.\ (\ref{E_m_2}) for
given dependencies $\alpha_{n,p}(E)$ makes it possible to determine $E_{\rm
m}$ as a function of $v_f$ and $\sigma_0/N_d$ for any
semiconductor material.

For the special cases (a,b,c,d,e) (see Sec.\ 2E) and the Townsend's dependence
$\alpha(E)=\alpha_0 \, \exp(-{E_0}/{E})$ Eq.\ (\ref{E_m_2}) yields
\begin{eqnarray}
\label{E_m_Sym}
\int_0^{E_{\rm m}/E_0} \exp\left(-\frac{1}{y}\right) d y = \frac{E^{\star}(v_f/v_s, \sigma_0/N_d, \alpha_0)}{E_0},\; \;
\\
\nonumber 
 \;E^{\star} \equiv \frac{b}{\alpha_0}  \frac{v_f ^2-v_s^2}{2 v_f  v_{s}}
\left[ \ln \left(\frac{v_f}{v_s} \frac{N_d}{\sigma_0}\right)- 1 \right] \qquad \qquad \; \; \\
\label{E_m_a}
\qquad \qquad \qquad \qquad \qquad \qquad {\rm for~case~(a)},\\
\nonumber
 \; E^{\star} \equiv \frac{b}{\alpha_0}  \frac{v_f \mp v_s}{v_{s}}
\left[ \ln \left(\frac{v_f}{v_s} \frac{N_d}{\sigma_0}\right)- 1 \right]\qquad \qquad \; \; \\
\label{E_m_bc}
\qquad \qquad \qquad \qquad \qquad \qquad {\rm for~cases~(b,c)},
\\
\nonumber
 \; E^{\star} \equiv \frac{b}{\alpha_0}  \frac{v_f \mp v_s}{v_{s}}
\left[\ln \left(\frac{2v_f \mp v_s}{v_{s}}\frac{N_d}{\sigma_0}\right)-1\right] \qquad 
\\ \label{E_m_de}
\qquad \qquad \qquad \qquad \qquad \qquad {\rm for~cases~(d,e)}.
\end{eqnarray}
The dimensional coefficient $b/\alpha_0$ in $E^{\star}$ has a
meaning of change of the electric field on the length of impact
ionization $\alpha_0^{-1}$ for the slope $d_z E=b$ which is
determined by the doping level $N_d$. For realistic parameters
$b/\alpha_0 \ll E_0$. Comparing  Eq.\ (\ref{E_m_bc}) with Eq.\
(\ref{E_m_de}) we see that asymmetry of transport properties has
logarithmically weak effect on $E_{\rm m}$.

We solve Eq.\ (\ref{E_m_Sym}) numerically and
show $E_{\rm m}/E^{\star}$ as a function of $E_0/E^{\star}$ in
Fig.\ \ref{Figfield1}.
Together with Eqs.\ (\ref{E_m_a},\ref{E_m_bc},\ref{E_m_de})
this dependence makes it possible to determine $E_{\rm m}$
for given values of $v_f/v_s$, $\sigma_0/N_d$,
$\alpha_0$ and $E_0$ for the special cases (a--e).
[Note that $E_{\rm m}/E^{\star} \rightarrow 1$
when $E_0/E^{\star} \rightarrow 0$. This limit corresponds to
ultrastrong electric field when $\alpha(E) \rightarrow \alpha_0$.]
$E_{\rm m}$ increases with $E_0$
because the effective threshold of impact ionization becomes higher.

In Fig.\ \ref{Figfield2} we show $E_{\rm m}$ as a function of
$v_f/v_s$ for different values of $\sigma_0/N_d$. Thick solid
lines 1a, 2a, 3a correspond to the symmetric case (a) [Eq.\
(\ref{E_m_a})], thin solid lines 1b, 2b, 3b and dashed lines 1c,
2c, 3c correspond to the cases  of ``monopolar ionization" (b) and
(c) [Eq.\ (\ref{E_m_bc})], respectively. $E_{\rm m}$ is the
smallest for the symmetric case $\alpha_n=\alpha_p$ because both
types of carriers are involved in impact ionization. Curves 1c,
2c, 3c are ``discontinuous": $E_{\rm m}$ tends to a finite value
when $v_f/v_s \rightarrow 1$ whereas we expect $E_{\rm m}=0$
for $v_f/v_s=1$ as it is for the curves 1a, 2a, 3a and 1b, 2b, 3b.
This feature results from the assumption $\alpha_n = 0$ and
disappears for arbitrary small but nonzero value of $\alpha_n$.
Physically it means that impact ionization by electrons that move
parallel to the front is important at low front velocities $v_f
\sim v_s$ even in the case $\alpha_n \ll \alpha_p$. The curves for two
types of ``monopolar ionization'' (b) and (c) become close with
increase of $v_f$. The dependencies $E_{\rm}(v_f)$ calculated for
fully immobile holes or electrons [cases (d) and (e), Eq.\
(\ref{E_m_de}), the respective curves not shown]  turn out to be
very close to the cases of monopolar ionization (b) and (c) [Eq.\
(\ref{E_m_bc})], curves 1b, 2b, 3b and 1c, 2c, 3c, respectively]:
the difference  is smaller than 2\% in the whole interval
$1<v_f/v_s<100$. It means that $E_{\rm m}$ is much more influenced
by the asymmetry of impact ionization coefficients than by the
asymmetry of drift velocities. We also see that $E_{\rm m}$ and
hence the effective width of ionization zone $\ell_f$ decrease
with $\sigma_0$. This observation explains why the concentration
$\sigma^{\star}$ (see Fig.\ \ref{sigmaStar}) also decreases with
$\sigma_0$. In Fig.\ \ref{Figfield2}  we choose $b/ (\alpha_0 E_0)
= 0.0002$. For Si this value corresponds to $N_d=10^{14} \; {\rm
cm^{-3}}$ which is a typical doping level for high voltage  Si
structures. \cite{GRE79,GRE81,GRE81a} Since in Si impact
ionization by electrons dominates and $v_{ns} \approx v_{ps}$
(Ref.\ \onlinecite{handbook}), curves 1b, 2b, 3b provide a good
approximation for this material.

Fig.\ \ref{Figfield3} shows the dependence $E_{\rm m}(v_f)$ for
different values of $b/(\alpha_0 E_0)$ and the fixed value of
$\sigma_0 / N_d$. $E_{\rm m}$ increases with $N_d$ due to the
decrease of the effective width of ionization zone $\ell_f$ and
decreases with $\alpha_0$ due to the more efficient impact
ionization.

The $E_{\rm m}(v_f)$ dependencies obtained for the case of
symmetric ionization $\alpha_n = \alpha_p$ and ionization by
electrons (curves 1a, 2a, 3a and 1b, 2b, 3b in Figs.\
\ref{Figfield2},\ref{Figfield3}) can be fitted by the squareroot
function
\begin{equation}
\label{squareroot}
E_{\rm m}(v_f) - E_{\rm th} \sim E_0 \sqrt{(v_f / v_s) - 1},
\end{equation}
where $E_{\rm th} \approx 0.15...0.2 \, E_0$ plays
the role of the effective threshold of
impact ionization. This fit is quantitatively accurate for $v_f/v_s > 2$
and remains qualitatively correct for $1 < v_f/v_s < 2$.
Straightforward examination of Eq.\ (\ref{E_m_2}) shows that
the squareroot dependence corresponds to the piece-wise linear
approximation of the impact ionization coefficient $\alpha(E)$
\begin{equation}
\label{TownsendApproximation}
\alpha(E) \sim (E-E_{\rm th}) \, \Theta(E-E_{\rm th}),
\end{equation}
where $\Theta(E)$ is the step function. The function (\ref{TownsendApproximation})
approximates the Townsend's dependence
$\alpha(E)= E_0 \, \exp(-E_0/E)$ reasonably well in the most important
interval of electric fields $0.3...0.7 \, E_0$
(see inset to Fig.\ \ref{Figfield3}).
[Previously the approximation (\ref{TownsendApproximation})
has been discussed in the theory of finger-like streamers in Ref.\ \onlinecite {Raizer}.]
It is remarkable that the dynamics of ionization fronts reveals the existence
of effective threshold electric field $E_{\rm th} \approx 0.2 \, E_0$
in spite of the absence of any kind of cut-off at low
electric fields in the Townsend's dependence itself.
In particular, the squareroot dependence (\ref{squareroot})
implies that $v_f \sim {\ell_f}^2$ if we define
$\ell_f \approx (E_{\rm m}- E_{\rm th})/b$.

\subsection{The width of the screening region}

The screening region is situated just behind the ionization zone
(Fig.\ \ref{sketch1}).
Here the electric field is insufficient for impact ionization, but
the drift velocities remain saturated. According to
Eqs. (\ref{sigma4},\ref{current4}) in this interval of electric fields
the concentration $\sigma$ and space charge $\rho$ are conserved and
keep values $\sigma^{\star}$ and $\rho^{\star}$ determined by
Eqs.\ (\ref{sigma_star},\ref{rho_star}).
The slope of electric field in this region is determined as
$$
d_z E = \frac{q}{\varepsilon \varepsilon_0} (\rho^{\star}+N_d).
$$
Taking into account
Eqs.\ (\ref{sigma_m},\ref{sigma_star})
we find the ratio between the slope $|d_z E|$ in the screening region
and the slope $b$ in the depleted region
\begin{equation}
\label{slope}
\frac{|d_z E|}{b}=
\frac{v^{+}_{s}}{v_f+v^{-}_{s}} \, \frac{\sigma^{\star}}{N_d}-1=
\ln \frac{\sigma^{\star}}{\sigma_0}-1.
\end{equation}
The width $\ell_{\rho}$ of the screening region can be
evaluated as
\begin{equation}
\label{ell_rho1}
\ell_{\rho} \approx \frac{E_{\rm m}}{|d_z E|}
=\frac{E_{\rm m}}{b[\ln(\sigma^{\star}/\sigma_0)-1]}.
\end{equation}
Calculating $\ell_{\rho}$ in this way we include in it a part of ionization zone
(see Fig.\ \ref{sketch1}) and hence somewhat
overestimate the actual width of the screening region.
On the other hand, Eq.\ (\ref{ell_rho1}) gives an idea of the effective front
width since it accounts for the two regions of most dramatic change
in concentration and electric field:
the region of most rapid increase
of concentration from $\sigma_{\rm m}$ to the final value $\sigma^{\star}$
(approximately by one order of magnitude) and the region
of steep drop of electric field from $E_{\rm m}$ to $E_{\rm pl} \ll E_{\rm m}$.

We show $|d_z E|/b$ as a function of $v_f/v^{+}_s$ in Fig.\
\ref{fieldSlope}. Solid lines correspond to the symmetric case
$v^{-}_s=0$ for different values of $\sigma_0$. These curves
correspond also to the special cases (a,b,c) if $v^{+}_s$ is
replaced by $v_s$. The electric field profile  in the screening
region is approximately 10...20 times steeper than in depleted
region. The slope $|d_z E|/b$ has weak logarithmic dependence on
$v_f$ and decreases with $\sigma_0/N_d$. Dashed lines and dotted
lines represent $|d_z E|/b$ for two limiting asymmetric cases
$v^{-}_s/v^{+}_s=-1$ [case (d),immobile holes] and
$v^{-}_s/v^{+}_s=1$ [case (e), immobile electrons]. Similar to
$\sigma^{\star}$, $|d_z E|/b$ has a weak dependence on the
asymmetry of saturated drift velocities ($v^{-}_s \ne 0$) for
sufficiently large $v_f$.

\subsection{Transition from high-field to low-field region}

In the region where electric field is insufficient for impact ionization
it follows from Eq.\ (\ref{sigma4}) that
\begin{equation}
\label{sigma5}
\frac{[v_{f}+v^{-}(E)]^2-[v^{+}(E)]^2}{v_f+v^{-}(E)} \, \sigma +
\frac{v^{+}(E)}{v_f+v^{-}} \, j
= {\rm const},
\end{equation}
where the  second small term is negligible. Employing conservation
of this quantity, we find explicit expressions
for  $\sigma$ and $\rho$:
\begin{eqnarray}
\label{sigma5a}
\sigma(E)= \frac{[v_f^2+v^{-}_{s}]^2-[v^{+}_s]^2}{[v_f^2+v^{-}(E)]^2-[v^{+}(E)]^2} \,
\frac{v_f+v^{-}(E)}{v_f+v^{-}_{s}}\, \sigma^{\star},&&\\ \nonumber
\rho(E) = -\frac{v^{+}(E)}{v_f+v^{-}_{s}} \,
\frac{[v_f^2+v^{-}_{s}]^2-[v^{+}(E)]^2}{[v_f^2+v^{-}(E)]^2-[v^{+}(E)]^2}\,
\sigma^{\star}. \; \; \; &&
\end{eqnarray}
For the special cases (a,b,c) this yields
\begin{eqnarray}
\label{sigma5b}
\sigma(E)&=& \frac{v_f^2-v_s^2}{v_f^2-v^2(E)}\, \sigma^{\star}; \\ \nonumber
\rho(E) &=& - \frac{v(E)}{v_f} \, \frac{v_f^2-v_s^2}{v_f^2-v^2(E)} \, \sigma^{\star}.
\end{eqnarray}
If $v(E)$ is monotonic, then according to Eq.\ (\ref{sigma5b})
$|\rho(E)|$ monotonically decreases with decrease of $E$. The
transition to neutral plasma occurs when $|\rho(E)|$ reaches $N_d$
and $E$ reaches a certain constant asymptotic value. In
semiconductors with nonmonotonic $v_n(E)$ dependence (e.g., GaAs)
that has maximum at $\widetilde E < E_s$, we expect that
$|\rho(E)|>|\rho^{\star}|$ near $\widetilde E$, but the transition
to plasma at lower electric fields occurs in the same way as for
monotonic $v(E)$.

\subsection{Parameters of the plasma region}

Plasma concentration and electric field in plasma are denoted
as $\sigma_{\rm pl}$ and $E_{\rm pl}$, respectively (Fig.\ \ref{sketch1}).
Generally, these parameters are determined by Eqs.\ (\ref{current4},\ref{sigma5})
together with the neutrality condition $\rho_{\rm pl}=-N_d$.

For the general asymmetric case $v^{-}(E) \ne 0$
we approximate the drift velocities in plasma by
the Ohm low $v_n(E)=\mu_n \, E$, $v_p(E)= \mu_p \, E$ (note that
for the approximations (\ref{velocityDrift}) $\mu_{n,p}=v_{ns,ps}/E_{ns,ps}$).
This yields
\begin{eqnarray}
\label{sigma6}
&&\frac{(v_f+v^{-}_{s})^2-(v^{+}_{s})^2}{v_f+v_{-s}} \, \sigma^{\star} =\\
\nonumber
&& \qquad \qquad \qquad \qquad 
=\frac{(v_f+\mu^{-} E_{\rm pl})^2-(\mu^{+} E_{\rm pl})^2}{v_f+\mu^{-}E_{\rm pl}} \, \sigma_{\rm pl}, 
\\ \label{current6}
&&j=\mu^{+} \sigma_{\rm pl} E_{\rm pl} + (v_f+\mu^{-}E_{\rm pl})(-N_d), \\ \nonumber
&&\qquad \qquad \qquad \qquad \qquad \qquad \qquad 
\mu^{\pm}=\frac{\mu_p \pm \mu_n}{2}.  
\end{eqnarray}
Expressing $E_{\rm pl}$ via $\sigma_{\rm pl}$ in Eq.\
(\ref{current6}), neglecting $j$ and substituting in Eq.\
(\ref{sigma6}) we find explicit formulas for $\sigma_{\rm pl}$
\begin{eqnarray}
\label{sigmaPlasma}
&&\sigma_{\rm pl}= \frac{1}{2}A\,\sigma^{\star}\left[1 +
\sqrt{1+\frac{4 \,N_d}{A \sigma^{\star}}\left( \frac{N_d}{A \sigma^{\star}}-
\frac{\mu^{-}}{\mu^{+}}\right)}\right], \\ \nonumber
&& \qquad \qquad \qquad \qquad \qquad A \equiv \frac{(v_f+v^{-}_{s})^2-(v^{+}_{s})^2}{v_f \,(v_f+v^{-}_{s})},
\\ \label{fieldPlasma}
&& E_{\rm pl}=E^{+}_s \, \frac{v_f / v^{+}_s}{\sigma_{\rm pl}/N_d - \mu^{-}/\mu^{+}},
\qquad E^{+}_s \equiv \frac{v^{+}_s}{\mu^{+}}.
\end{eqnarray}
Expansion  over $N_d/A \sigma^{\star}$ leads to
\begin{eqnarray}
\label{sigmaPlasma1}
&&\sigma_{\rm pl}=A\, \sigma_{\star}+\frac{N_d^2}{A \sigma^{\star}}-\frac{\mu^{-}}{\mu^{+}}\,N_d \approx
\\ \nonumber && \qquad \qquad \qquad \qquad \approx 
A\, \sigma^{\star}=\frac{(v_f+v^{-}_{s})^2-(v^{+}_{s})^2}{v_f \,(v_f+v^{-}_{s})} \,  \sigma^{\star}.
\end{eqnarray}
Electron and hole concentrations in plasma region are recovered as
\begin{equation}
\label{npPlasma}
n_{\rm pl}= \frac{\sigma_{\rm pl} + N_d}{2}, \qquad p_{\rm pl}= \frac{\sigma_{\rm pl} - N_d}{2}.
\end{equation}
It can be easily derived from the continuity of electron and hole flows that
$n^{\star} > n_{\rm pl}$ and $p^{\star} < n_{\rm pl}$ [see Eq.\ (\ref{np_star})].
In contrast, the relation between
$\sigma^{\star}$ and $\sigma_{\rm pl}$ is not universal: $A<1$ when $v_s^{-} \le 0$,
but $A >1$ for $v^{-}_s > 0$ and $v_f > \left[(v_s^{+})^2-(v_s^{-})^2\right]/(v_s^{-})^2$.
Therefore according to Eq.\ (\ref{sigmaPlasma1})
$\sigma_{\rm pl} < \sigma^{\star}$ when $v_{ns}>v_{ps}$ for any $v_f$.
However, for sufficiently large $v_f$ we find that $\sigma_{\rm pl} > \sigma^{\star}$
when $v_{ns} < v_{ps}$. Next, $A \rightarrow 0$ for $v_f \rightarrow v_{ns}$ and hence
according to Eqs.\ (\ref{sigmaPlasma},\ref{npPlasma})
$\sigma_{\rm pl} \rightarrow N_d$, $n_{\rm pl} \rightarrow N_d$,
$p_{\rm pl} \rightarrow 0$. This is consistent with condition $v_f > v_{ns}$.

For the  special cases (a,b,c) when electron and hole drift
velocities are equal it is convenient to use the approximation
$v(E)= v_s E/ (E + E_{\rm s})$ in the whole range of electric
fields including plasma. After similar derivations we find
\begin{equation}
\label{plasmaSym1}
\sigma_{\rm pl}=\sigma^{\star} \cdot \frac{v_f^2-v_s^2}{2 v_f^2} \cdot
\left[1+\sqrt{1+\left(\frac{2 v_f^2}{v_f^2-v_s^2} \frac{N_d}{\sigma^{\star}} \right)^2}  \right],
\end{equation}
\begin{eqnarray}
&&E_{\rm pl} = E_s \, \frac{D}{1-D}, \\ \nonumber 
&&D \equiv \frac{v_f^2-v_s^2}{2 v_s^2} \cdot
\left[ \sqrt{1+\left(\frac{2 v_f^2}{v_f^2 - v_s^2} \frac{N_d}{\sigma^{\star}}\right)^2}-1 \right]
\cdot
\ln \frac{\sigma^{\star}}{\sigma_0}.
\end{eqnarray}
Expanding over $N_d/\sigma^{\star}$, we obtain simplified
dependencies that are valid when $\sigma^{\star} \gg N_d$
\begin{eqnarray}
\label{plasmaSym2}
\sigma_{\rm pl}=\frac{v_f^2-v_s^2}{v_f^2} \,
\left(\sigma^{\star} + \frac{v_f^2}{v_f^2-v_s^2} \frac{N_d^2}{\sigma^{\star}} \right)
\approx \frac{v_f^2-v_s^2}{v_f^2} \cdot \sigma^{\star}, &&
\\ \nonumber
E_{\rm pl}={E_s} \left[ \frac{v_f^2-v_s^2}{v_f^2} \,
\ln \frac{\sigma^{\star}}{\sigma_0} - 1 \right]^{-1}. \qquad \qquad \qquad \; &&
\end{eqnarray}
The concentration $\sigma_{\rm pl}$ tends to $\sigma^{\star}$ with increase of
$v_f$ and, similar to $\sigma^{\star}$, shows quasilinear dependence on $v_f$ for $v_f \gg v_s$.
In contrast, the dependence of  $E_{\rm pl}$ on $v_f$ is very weak.

The concentration $\sigma_{\rm pl}$ and the electric field $E_{\rm
pl}$ are shown in Figs. \ref{FigSigmaPlasma} and
\ref{FigFieldPlasma}, respectively. In panels (a) these
dependencies are shown for different values of $\sigma_0/N_d$.
Solid curves 1, 2, 3 correspond to the symmetric case $v^{-}_s=0$.
Dashed curves 1d, 2d, 3d correspond to the limiting case (d) of
immobile holes $v^{-}_s/v^{+}_s=-1$. Dotted curves 1e, 2e, 3e
correspond to the opposite limiting case (e) of immobile electrons
$v^{-}_s/v^{+}_s=1$. For fast fronts $\sigma_{\rm pl}$ increases
with $v_f$ linearly, whereas the electric field is close to
$E_{\rm pl} \approx 0.1 \, E_s^{+}$ and weakly decreases with
$v_f$. In panels (b) $\sigma_{\rm pl}$ and  $E_{\rm pl}$ are shown
for different values of $v^{-}_s/v^{+}_s$ and $\mu^{-}/\mu^{+}$.
We see that the asymmetry in high-field transport is much more
important than the asymmetry in the low-field transport. For
$v_s^{-}>0$ (curves 4,5,6,7 in Fig.\ \ref{FigFieldPlasma}(b)) the
dependence $\sigma_{\rm pl}(v_f)$ has a kind of plateau and the
corresponding dependence $E_{\rm pl}(v_f)$ has minimum. This
occurs in the interval of front velocities $v_{ns} < v_f <
v_{ps}$. The plateau on $\sigma_{\rm pl}(v_f)$ is caused by the
peculiarity of $\sigma^{\star}(v_f)$ dependence that has been
discussed above in Sec.\ 3C (see also Fig.\ \ref{sigmaStar}). The
nonmonotonic behaviour of $E_{\rm pl}(v_f)$ becomes clear if we
take into account that according to Eq.\ (\ref{fieldPlasma}) in
symmetric case $E_{\rm pl} \sim v_f/ \sigma_{\rm pl}$ : $E_{\rm
pl}$ decreases and increases when the increase of $\sigma_{\rm pl}$
with $v_f$ is superlinear and sublinear, respectively. Hence
``plateau" on the dependence $\sigma_{\rm pl}(v_f)$ corresponds to
minimum on $E_{\rm m}(v_f)$.

\subsection{Voltage over the structure}

The voltage over the structure is given by the integral
\begin{equation}
\label{voltage1}
u=\int_0^W E(x) dx.
\end{equation}
We approximate the actual profile $E(z)$ by a piece-wise linear
profile A--B--C--D shown in Fig.\ \ref{sketch3} neglecting the voltage
drop over the plasma region where the electric field is low.
For such profile $\sigma(z)=\sigma_0$ and $d_z E= b - c \sigma_0$ on
part A--B, $E=E_{\rm m}$ and $\sigma$ increases from
$\sigma_0$ to $\sigma^{\star}$ in arbitrary way  on part B--C,
and $\sigma(z)=\sigma^{\star}$ and $d_z E = b - c\sigma^{\star}$ on part C--D.
Integration over this profile gives an upper bound for the actual voltage.
On the $(\sigma,E)$ plane this profile corresponds to the rectangular
$\sigma(E)$ dependence (dashed line A--B--C--D in Fig.\ \ref{sketch2}).
The integral over $z$ can be replaced either by the integral over
electric field $E$ or over  concentration $\sigma$
since according to Eqs.\ (\ref{sigmaHigh},\ref{fieldHigh})
$dz ={d E}/{(b - c \, \sigma)}= {d \sigma}/{ \lambda \, \beta_{\rm eff}\, \sigma}$.
Employing integration over $E$ for branches A--B and C--D
and integration over $\sigma$ for branch B--C , we approximate
the integral (\ref{voltage1}) as
\begin{equation}
\label{voltage2}
u=\int_{E_{\rm left}}^{E_{\rm m}} \frac{E \, dE}{b-c\sigma_0}
\qquad +\int_{\sigma_0}^{\sigma^{\star}} \frac{E_{\rm m} \,d \sigma}
{\lambda \, \beta_{\rm eff}(E_{\rm m}) \, \sigma}+
\int_{E_{\rm m}}^0 \frac{E \, dE}{b -c \sigma^{\star}} \, .
\end{equation}
This yields
\begin{eqnarray}
\label{voltage3}
&&u \approx   \frac{1}{2c} \left[ \frac{E_{\rm m}^2-E_{\rm left}^2}{\sigma_{\rm m}-\sigma_0}+
\frac{E_{\rm m}^2}{\sigma^{\star}-\sigma_{\rm m}} \right]+ \\
\nonumber && \qquad \qquad \qquad \qquad \qquad \qquad \qquad \;
+\frac{E_{\rm m}}{\lambda \, \beta_{\rm eff}(E_{\rm m})}
\ln \frac{\sigma^{\star}}{\sigma_0} \,.
\end{eqnarray}
Here  the first term corresponds to the contribution
of the inclined parts A--B and C--D of the field profile.
The second term corresponds to the horizontal part B--C
and hence approximates the contribution of the nonlinear part
of the profile near its maximum. The ratio of the first term
to the second one can be estimated as $v_s \, \alpha(E_{\rm m}) \, \tau_{\rho}$,
where $v_s \, \alpha(E_{\rm m})$ is the frequency
of impact ionization, $\tau_{\rho} \equiv \ell_{\rho}/v_f$ is
the time the front takes to move over its own width $\ell_{\rho}$
and it is assumed that $v_n=v_p$, $\alpha_n=\alpha_p$, $v_f \gg v_s$.
[For this estimate we assume $E_{\rm m} \gg E_{\rm left}$ and employ
Eq.\ (\ref{ell_rho1}).]
Typically $v_s \, \alpha(E_{\rm m}) \, \tau_{\rho} \gg 1$ and hence
the first term in Eq.\ (\ref{voltage3}) dominates over the second one.
It means that the nonlinear part of the profile $E(z)$ which is located
near its maximum is small, and the typical shape of $E(z)$ is close to triangle.

Taking into account that $E_{\rm left}=E_{\rm m} - b \, x_f$, we
can represent $u$ as
\begin{eqnarray}
\label{voltage4}
u \approx E_{\rm m} \, x_f - \frac{b \, x_f^2}{2} +
\frac{E_{\rm m}^2}{2b \,[\ln(\sigma^{\star}/\sigma_0)-1]}\, + \qquad \qquad && \\ \nonumber
+\frac{E_{\rm m}}{\lambda \, \beta_{\rm eff}(E_{\rm m})}
\ln \frac{\sigma^{\star}}{\sigma_0}. &&
\end{eqnarray}
Since the dependencies $\sigma^{\star}(v_f)$ and $E_{\rm}(v_f)$
are already determined [ see Eqs.\
(\ref{sigma_star},\ref{E_m_2})], Eq.\ (\ref{voltage4}) gives $u$
as a function of front position $x_f$ and velocity $v_f$.

\section{Ultrafast fronts ($v_f \gg v^{+}_s$)}

The front velocity $v_f$ is  often much higher
than $v_s$. \cite{GRE79,GRE81,GRE81a,BEN85,ALF87,EFA88,EFA90}
In the respective limiting case $v_f/v^{+}_{s} \gg 1$
the effect of transport asymmetry vanishes.
As it follows from Eqs.\ (\ref{sigma_m},\ref{sigma_star},\ref{sigmaPlasma1})
the concentrations $\sigma_{\rm m}$ and $\sigma^{\star}$
can be presented as
\begin{equation}
\label{concentrationsFast}
\sigma_{\rm m}=\frac{v_f}{v^{+}_{s}} \, N_d, \qquad
\sigma^{\star}=\frac{v_f}{v^{+}_{s}} \, N_d \, \ln \frac{\sigma^{\star}}{\sigma_0},
\end{equation}
The plasma concentration and electric field in plasma are given by
\begin{equation}
\label{plasmaFast}
\sigma_{\rm pl}=\left[1-  \left(\frac{v^{+}_{s}}{v_{f}}\right)^2 \right]
\, \sigma^{\star}\approx \sigma^{\star},
\;
E_{\rm pl}=\frac{E^{+}_{s}}{\ln (\sigma^{*}/\sigma_0)}.
\end{equation}

The maximum field is determined by [see Eq.\ (\ref{E_m_1})]
\begin{eqnarray}
\label{fieldFast1}
\int_{0}^{E_{\rm m}} \left[ v_{ns}\alpha_n (E)+v_{ps} \alpha_p(E)\right] \, dE= \qquad \qquad \qquad&& 
\\ \nonumber
=\frac{q N_d v_f}{\varepsilon \varepsilon_0} \ln \frac{v_f \, N_d}{v^{+}_s \, \sigma_0}.&&
\end{eqnarray}

Let us compare these predictions  with the results of an elementary
model suggested for planar ionization front
in diode structures in Ref.\ \onlinecite{ROD06} on the basis of
the ideas developed for finger-like streamers in Ref.\ \onlinecite{DYA88,DYA89}.
In Ref.\ \onlinecite{ROD06} it is assumed $v_{n}=v_{p}$, $v_f \gg v_s$,
$\alpha_p=0$ and   $\alpha_n(E)=\alpha_0 \, \Theta(E-E_{\rm th})$, where
$\Theta(E)$ is the step function.
Under these assumptions the order of magnitude values of $v_f$
and $\sigma_{\rm pl}$ has been evaluated as
\begin{eqnarray}
\label{speedFast2}
&&v_f = \frac{\ell_f}{\tau}, \;
\tau \equiv \frac{1}{v_s \alpha_0}
\ln \frac{\sigma_{\rm pl}}{\sigma_0}, \;
\ell_f \equiv \frac{E_{\rm m}-E_{\rm th}}{q N_d/\varepsilon \varepsilon_0}, \\ \nonumber
&&\sigma_{\rm pl} = \frac{\alpha_0 \varepsilon \varepsilon_0 E_{\rm m}}{q},
\end{eqnarray}
where $\tau$
is the time it takes for the front to pass over the width of ionization zone $\ell_f$.

For $\alpha_n(E)=\alpha_0 \, \Theta(E-E_{\rm th})$, $\alpha_p(E)=0$
we obtain from Eq.\ (\ref{fieldFast1})
\begin{equation}
\label{speedFast1}
v_f = \frac{\ell_f}{\widetilde \tau},
\qquad {\widetilde \tau}=\frac{1}{v_s \, \alpha_0}
\ln \frac{\sigma_{\rm m}}{\sigma_0}.
\end{equation}
Then it follows from Eqs.\ (\ref{concentrationsFast}, \ref{plasmaFast}, \ref{speedFast1})
that
\begin{equation}
\label{concentrationFast2}
\sigma_{\rm pl}= \frac{\alpha_0 \varepsilon \varepsilon_0 (E_{\rm m} - E_{\rm th})}{q}
- \sigma_{\rm m} \ln \frac{\sigma^{\star}}{\sigma_{\rm m}} \approx
\frac{\alpha_0 \varepsilon \varepsilon_0 (E_{\rm m} - E_{\rm th})}{q}.
\end{equation}

Predictions for $v_f$ given by
Eqs.\ (\ref{speedFast2}) and (\ref{speedFast1}) differ in the definition of $\tau$.
However,  the relative difference
$(\tau - \widetilde \tau) / \tau = \ln (\sigma^{\star}/\sigma_{\rm m})/ \ln (\sigma^{\star}/\sigma_0)$
does not exceed 10 \%  (see Fig.\ \ref{sigmaStar}). Next,
according to Eq.\ (\ref{speedFast2})
$\sigma_{\rm pl}$ is proportional to $E_{\rm m}$ whereas more accurate
Eq.\ (\ref{concentrationFast2}) predicts proportionality to the difference
between $E_{\rm m}$ and $E_{\rm th}$.
Therefore Eq.\ (\ref{speedFast2}) overestimates $\sigma_{\rm pl}$. Still it gives
correct order of $\sigma_{\rm pl}$ since in practice
$E_{\rm th}$, $E_{\rm m}$ and $|E_{\rm th} - E_{\rm m}|$ are
of the same order of magnitude.

\section{Nonstationary propagation}

\subsection{The adiabatic condition}

The relations between $v_f$, $E_{\rm m}$, $\sigma_{\rm pl}$ and
$E_{\rm pl}$ obtained for $J ={\rm const}$ still hold when $J$
varies in time providing that this variation is slow in comparison
with inner relaxation times of the traveling front. These times
are the Maxwellian relaxation time in plasma behind the front
$\tau_{\rm M}$ and the time $\tau_{\rho} \equiv \ell_{\rho}/v_f$
it the front takes to move over the width of the screening
region $\ell_{\rho}$. Indeed, any change of the electric field (and
hence the current density) in the $n$ base  originates from changes of
electric charges in the highly doped $p^{+}$ and $n^{+}$ layers
that serve as effective electrodes (see Fig.\ \ref{sketch1}).
Further transfer of electric charge into the $n$ base occurs
through the plasma layer and is controlled by $\tau_{\rm M}$.
Redistribution of charges in the traveling screening region at the
plasma edge takes time $\tau_{\rho}$. Thus the times $\tau_M$ and
$\tau_{\rho}$ characterize how fast the front relaxes to the
steady profile that corresponds to the instant value of the
current density. Employing Eq.\ (\ref{ell_rho1}), assuming for
simplicity $v_s = v_{ns} = v_{ps}$, $\mu = \mu_n = \mu_p$ and
taking into account $v_s = \mu E_s$, we obtain
\begin{equation}
\label{adiabatic1}
\tau_{\rho} \equiv \frac{\ell_{\rho}}{v_f}=
\tau_{\rm M} \, \frac{E_{\rm m}}{E_s} \, \frac{\sigma_{\rm pl}}{\sigma^{\star}},
\qquad \tau_{\rm M} \equiv \frac{\varepsilon \varepsilon_0}{q \mu \sigma_{\rm pl}} \,.
\end{equation}
We see that $\tau_{\rho}$ is much larger than $\tau_{\rm M}$ and hence
it is the time $\tau_{\rho}$ that eventually controls the relaxation
of the front profile.
Consequently, the adiabatic condition for the variation of current $J$
can be presented as
\begin{equation}
\label{adiabatic2}
\tau_{\rho} \cdot \frac{d(\ln J)}{d t} \ll 1 \, .
\end{equation}
Below we show that this condition is typically met for the realistic operation
mode of high voltage diodes used as switches in pulse power applications.

\subsection{Coupling to the external circuit}

In practice the device is connected to the voltage source $U(t)$ via a load resistance
$R$. The current density $J$ and the voltage over the structure $u$
are related via Kirchhoff's equation
\begin{equation}
\label{Kirchhoff}
u(t) + R S J(t)= U(t) \, .
\end{equation}
In high voltage diodes used in pulse power applications the front
passage switches the structure from the nonconducting state to the
conducting state. At the moment $t=t_0$ when the front starts to
travel $u \approx U(t_0)$ and  $J \approx 0$. The switching time
is determined as $\Delta t = W / \langle v_f \rangle$, where $W$
is the $n$ base width and $\langle v_f \rangle$ is the mean value
of front velocity (generally, $v_f$ increases during the front
passage). The device resistivity after switching is negligible in
comparison with the load resistance $R$. Hence $u(t_0+\Delta t)
\approx 0$ and $J(t_0 + \Delta t) = U(t_0+\Delta t)/(R S) \approx
U(t_0+\Delta t)/(RS)$, where we take into account that variation
of $U$ within the time period $\Delta t$ is small. Then we
estimate the relative variation of the current density as $$
\frac{d(\ln J)}{d t} \sim \frac{J(t_0+\Delta t)-J(t_0)}{\Delta t
\, J(t_0+\Delta_t)} \sim \frac{1}{\Delta t} = \frac{W}{\langle v_f
\rangle}. $$ Substituting this estimate to Eq.\
(\ref{adiabatic2}), we present the adiabatic condition as
\begin{equation}
\label{adiabaticLast}
\frac{\langle v_f \rangle}{v_f} \, \frac{\ell_{\rho}}{W}
\approx \frac{\ell_{\rho}}{W} \ll 1.
\end{equation}
Eq.\ (\ref{adiabaticLast}) states that the inner dynamics of the
traveling front and the outer dynamics that is controlled by the
external circuit can be separated if the effective front width
$\ell_{\rho}$ is much smaller than the size of the system $W$.
Using Eq.\ (\ref{ell_rho1}) we present (\ref{adiabaticLast}) as
\begin{equation}
\label{adiabaticVeryLast}
\frac{E_{\rm m}}{b W} \, \frac{1}{\ln (\sigma^{\star}/\sigma_0)-1} \ll 1.
\end{equation}
According to Eq.\ (\ref{slope}) and Fig.\ (\ref{fieldSlope})
the second term in (\ref{adiabaticVeryLast}) has numerical value in the range
0.05...0.1. Therefore it is necessary that
$E_{\rm m}/(b W) \sim 1$.
The adiabatic conditions (\ref{adiabaticLast},\ref{adiabaticVeryLast}) are
met or nearly met for high-voltage sharpening diodes where
$W \sim 100...300 \, {\rm \mu m}$ and $\ell_{\rho} \sim 10...20 \; {\rm \mu m}$
but are not likely to be met for much smaller TRAPATT diodes.

In conclusion,
the relations between $v_f$, $E_{\rm m}$, $\sigma_{\rm pl}$ obtained for the self-similar
propagation mode can be used in general case to relate the instant values of these parameters
for sufficiently large structures and fast fronts.
In this case the voltage $u(v_f,x_f)$ given by Eq.\ (\ref{voltage4}) can
be substituted to the Kirchoff's equation (\ref{Kirchhoff}). Then
equation $d x_f/dt =- v_f$ (recall the $v_f >0$ for the front traveling in the negative $x$ direction)
together with Eq.\ (\ref{Kirchhoff})
represent a set of ordinary differential equations
that describe the front propagation with account taken for the external circuit.

\section{Summary}

\begin{figure*}[hp]
\begin{center}
\vskip 1cm
\includegraphics*[width=9.0 cm,height=11.5 cm,angle=270]{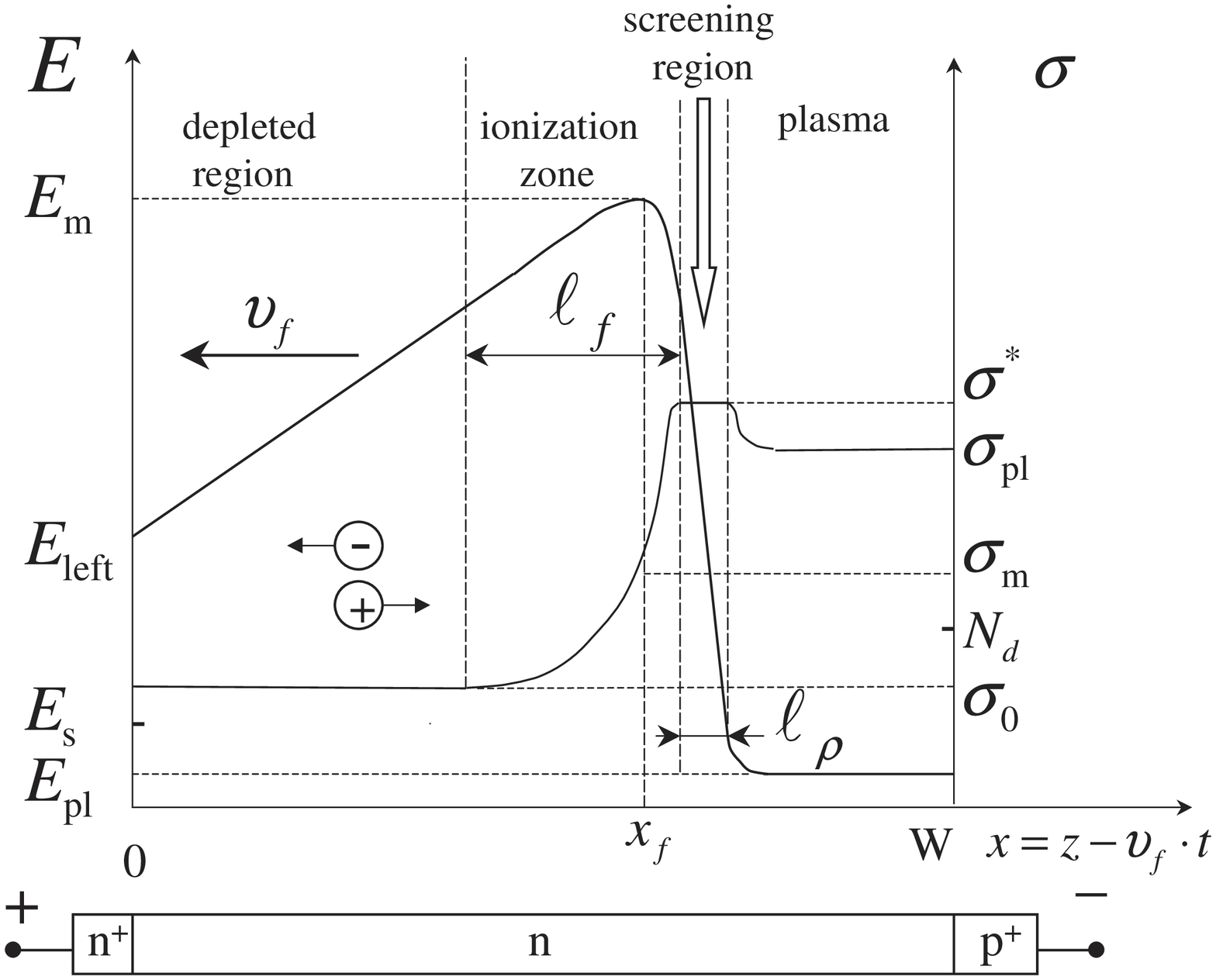}\\
\end{center}
\caption {Sketch of the electric field $E$ and total concentration
of free carriers concentrations $\sigma =n+p$ (lower panel) in the
$p^{+}$-$n$-$p^{+}$ structure during the passage of the ionization
front. The field $E_s$ corresponds to the transition from linear
low-field transport to saturated drift velocities. Coordinates $x$
and $z=x+v_f \, t$ correspond to stationary and comoving frames,
respectively. Note relations $\sigma_0 \ll
N_d,\sigma^{\star},\sigma_{\rm pl}$ and between the initial
concentration $\sigma_0$ in the depleted region, doping $N_d$ and
plasma concentration $\sigma_{\rm pl}$. The relation
$\sigma^{\star}>\sigma_{\rm pl}$ generally holds only for
$v_{ns}>v_{ps}$ and can be broken for $v_{ns} < v_{ps}$.}
\label{sketch1}
\end{figure*}

\begin{figure*}[hp]
\begin{center}
\includegraphics*[width=8.5 cm,height=12.0 cm,angle=270]{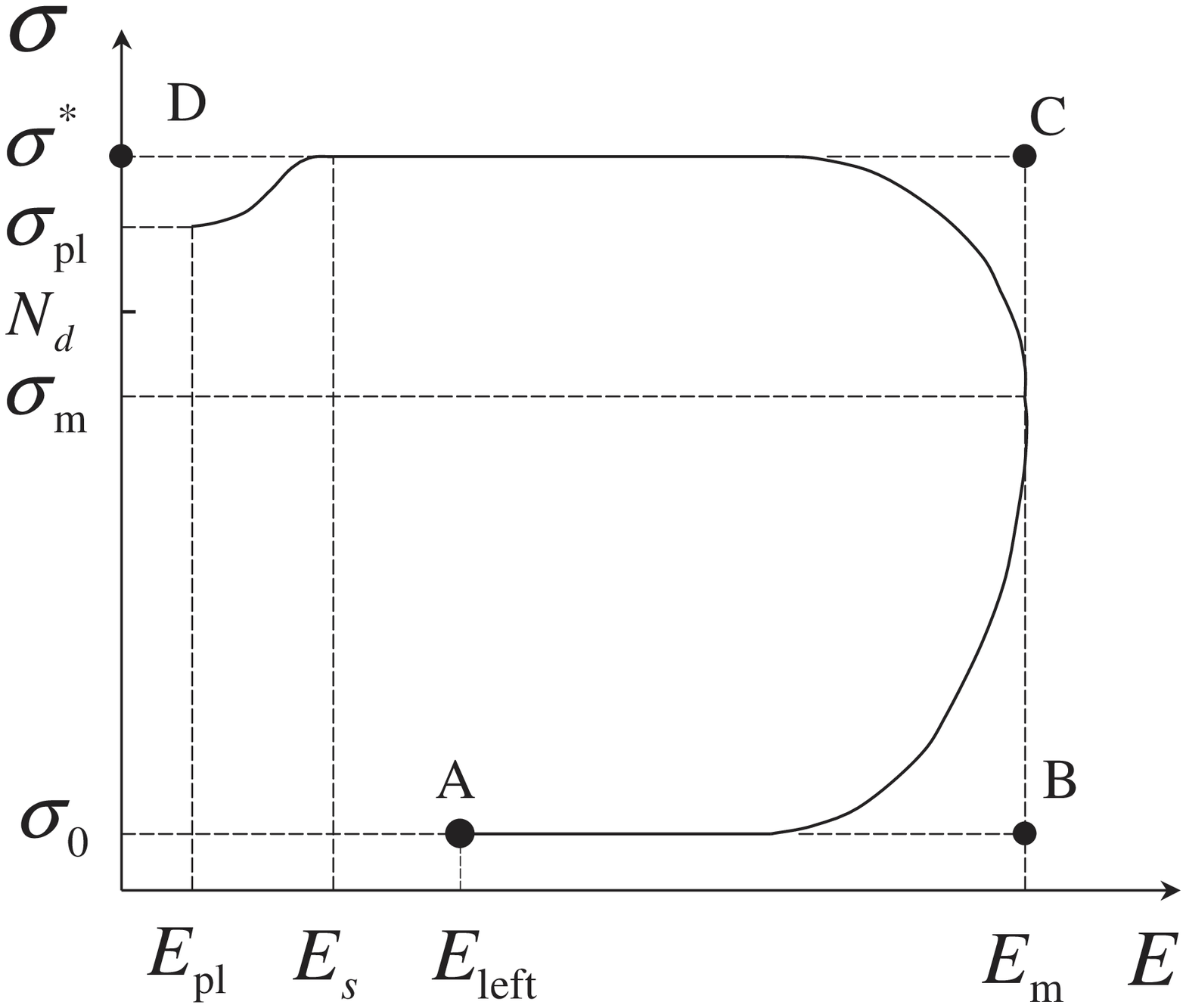}\\
\end{center}
\caption {Dependence of total carrier concentration on
electric field $\sigma (E)$ in the travelling ionization front.
See notations and comments to Fig.\ \ref{sketch1}. Path A--B--C--D
corresponds to piece-wise linear approximation of
the field profile shown in Fig.\ \ref{sketch3}.
}
\label{sketch2}
\end{figure*}

\begin{figure*}[hp]
\includegraphics*[width=8.5 cm,height=12.0 cm,angle=270]{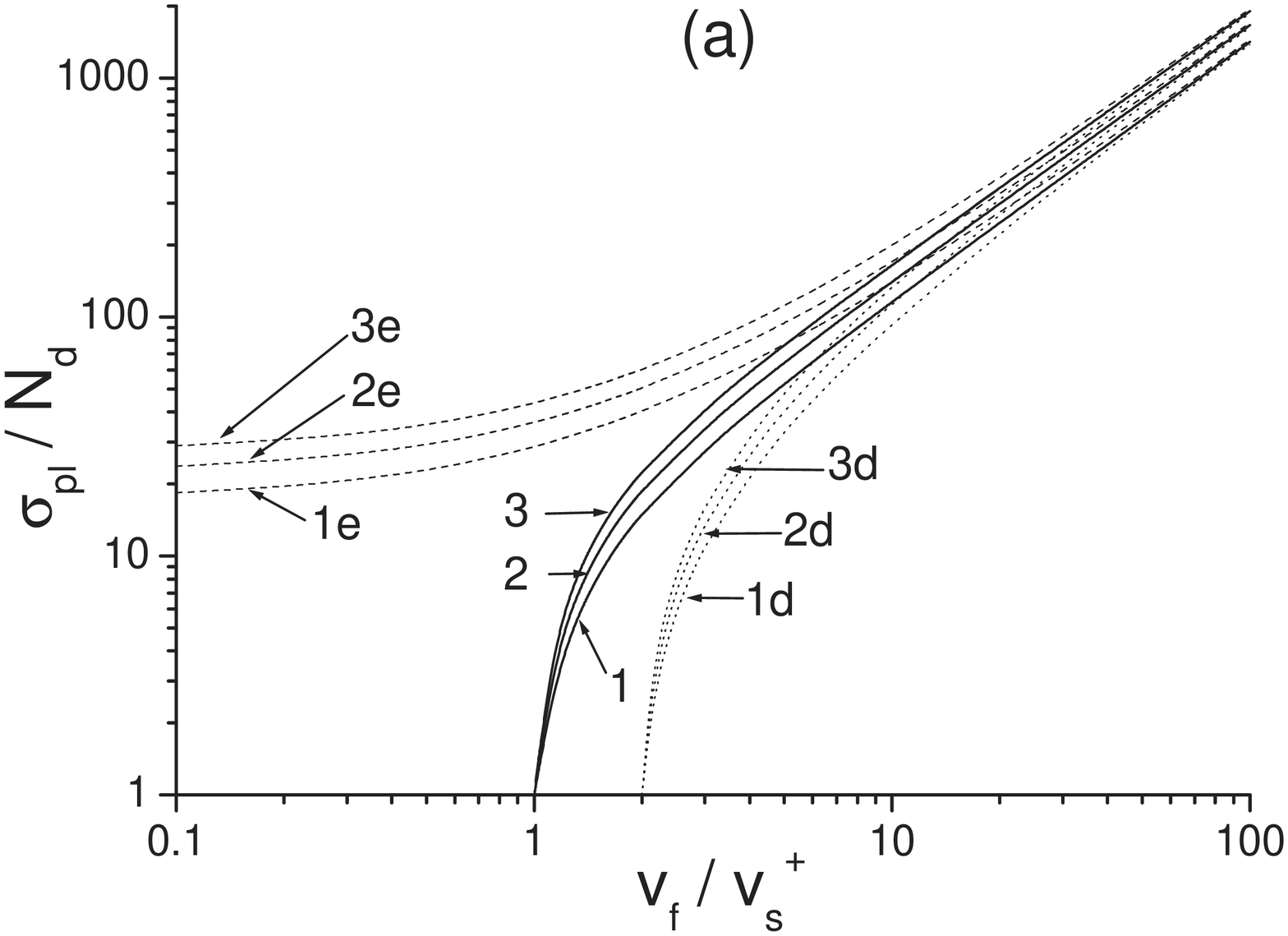}
\includegraphics*[width=8.5 cm,height=12.0 cm,angle=270]{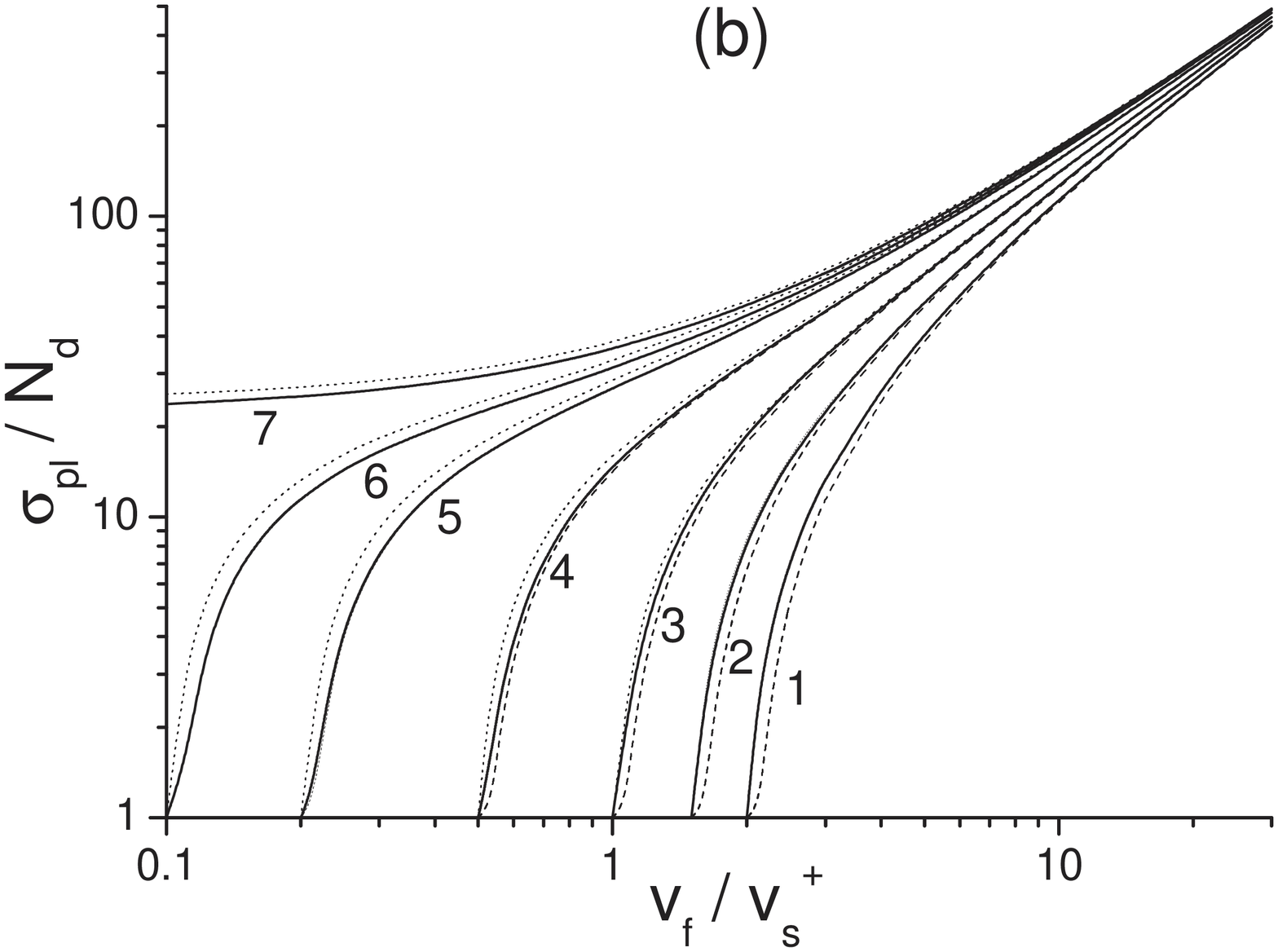}
\caption {Concentration of electron-hole plasma $\sigma_{\rm pl}$
generated by the front passage as a function of front velocity
$v_f$. In panel (a) the dependence $\sigma_{\rm pl}(v_f)$ is shown
for different values of $\sigma_0 / N_d$. Solid curves 1,2,3
correspond to the case of symmetric transport $v^{-}_s=0$,
$\mu^{-}=0$ (e.g. $v_{ns}=v_{ps}$, $\mu_n=\mu_p$). Dotted lines
1d,2d,3d and dashed lines 1e,2e,3e correspond to the limiting
cases of immobile holes $v_{ps}=0$, $\mu_p=0$ [case(d)] and
immobile electrons $v_{ns}=0$, $\mu_n=0$ [case (e)], and are
calculated for the same values of $\sigma_0 / N_d$. Curves of
1st,2nd and 3rd series correspond to $\sigma_0 / N_d =
10^{-3},10^{-4},10^{-5}$, respectively. In panel (b) the
dependence $\sigma_{\rm pl}(v_f)$ is shown for different values of
$v^{-}_s/v^{+}_s$ and $\mu^{-}/\mu^{+}$ and fixed value $\sigma_0
/ N_d = 10^{-4}$. Solid lines from 1 to 7 correspond to
$v^{-}_s/v^{+}_s=\mu^{-}/ \mu^{+}= -1,\, -0.5, 0,\,  0.5, \, 0.8,
0.9, 1.0$, respectively. Associated dotted and dashed lines in
panel (b) correspond to $\mu^{-}/\mu^{+}=-0.9$ and
$\mu^{-}/\mu^{+}=0.9$, respectively, and the same value of
$v^{-}_s/v^{+}_s$ as for the respective solid lines.}
\label{FigSigmaPlasma}
\end{figure*}

\begin{figure*}[hp]
\includegraphics*[width=8.5 cm,height=12.0 cm,angle=270]{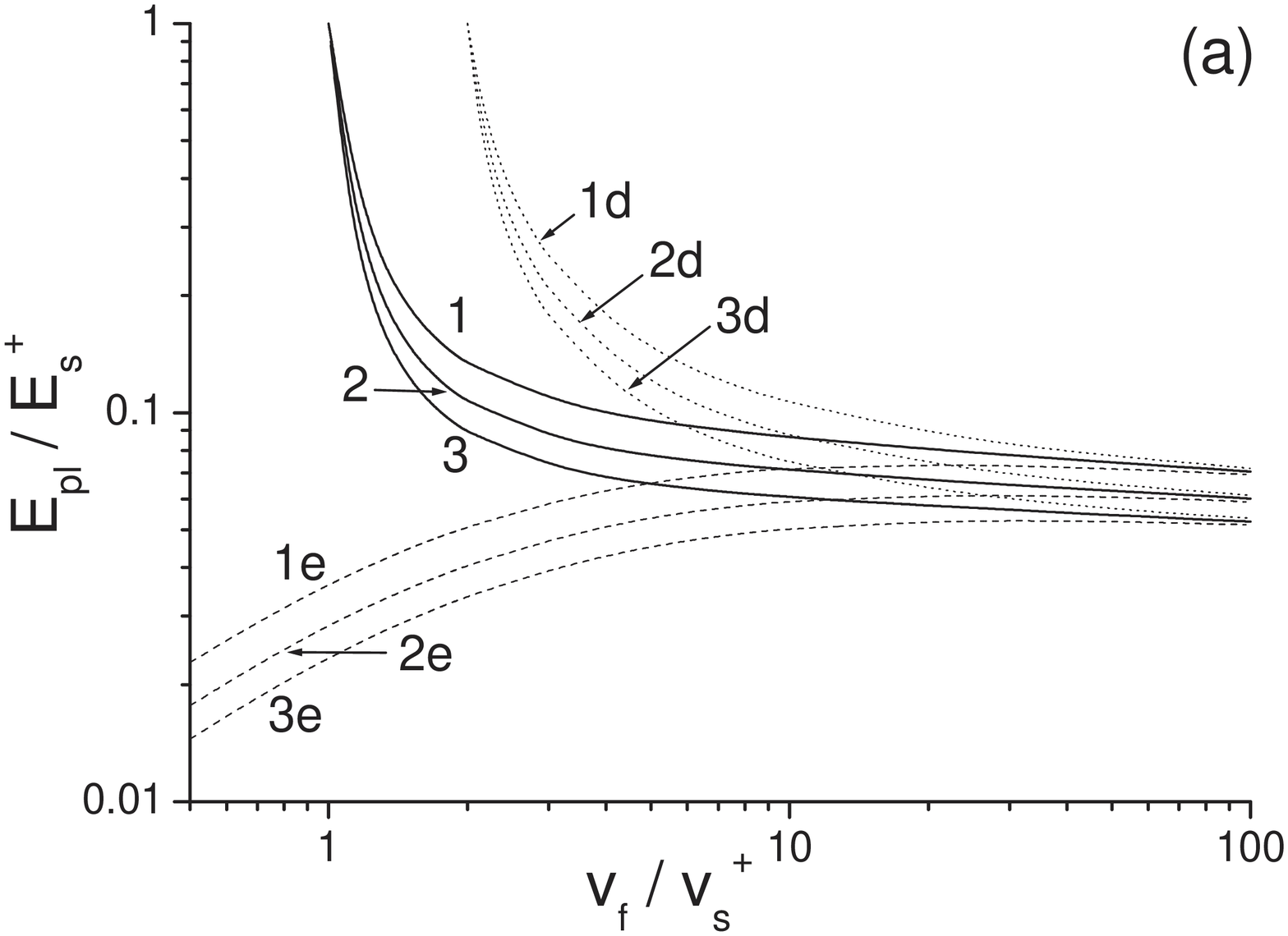}
\includegraphics*[width=8.5 cm,height=12.0 cm,angle=270]{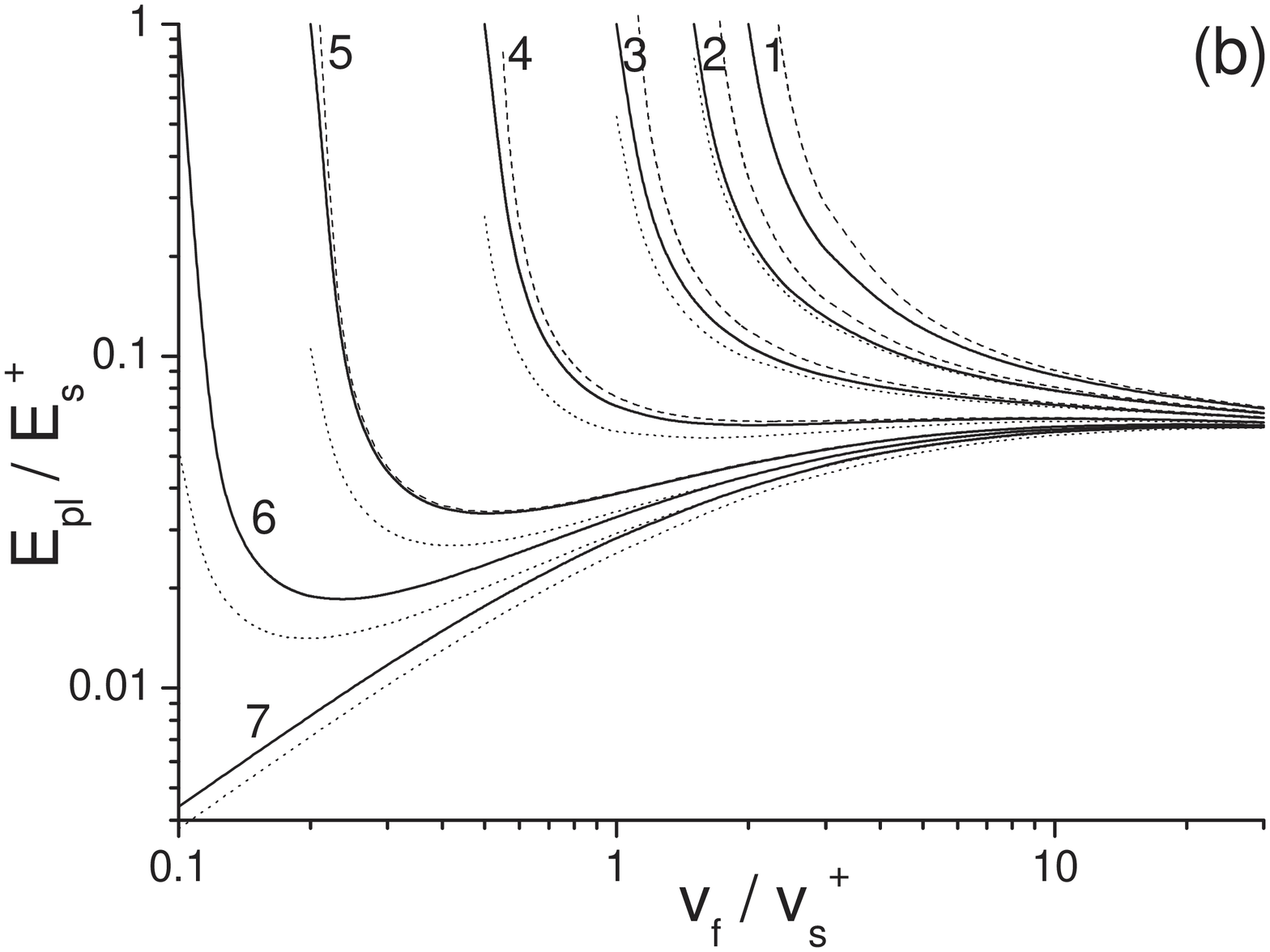}
\caption {Electric field  $E_{\rm pl}$ in the electron-hole plasma
generated by the front passage as a function of front velocity
$v_f$. In panel (a) the dependence $E_{\rm pl}(v_f)$ is shown for
different values of $\sigma_0 / N_d$. Solid curves 1,2,3
correspond to case of symmetric transport $v^{-}_s=0$, $\mu^{-}=0$
(e.g. $v_{ns}=v_{ps}$, $\mu_n=\mu_p$). Dotted lines 1d,2d,3d and
dashed lines 1e,2e,3e correspond to the limiting cases of immobile
holes  $v_{ps}=0$, $\mu_p=0$ [case(d)] and immobile electrons
$v_{ns}=0$, $\mu_n=0$ [case (e)], respectively. Curves of 1st,2nd
and 3rd series correspond to $\sigma_0 / N_d =
10^{-3},10^{-4},10^{-5}$, respectively. In panel (b) the
dependence $E_{\rm pl}(v_f)$ is shown for different values of
$v^{-}_s/v^{+}_s$ and $\mu^{-}/\mu^{+}$ and fixed value $\sigma_0
/ N_d = 10^{-4}$. Solid lines from 1 to 7 correspond to
$v^{-}_s/v^{+}_s=\mu^{-}/\mu^{+}= -1,\, -0.5, 0,\,  0.5, \, 0.8,
0.9, 1.0$, respectively. Associated dotted and dashed lines
correspond to $\mu^{-}/\mu^{+}=-0.9$ and $\mu^{-}/\mu^{+}=0.9$,
respectively, and the same value of $v^{-}_s/v^{+}_s$ as for the
respective solid lines.}
\label{FigFieldPlasma}
\end{figure*}

\begin{figure*}[hp]
\begin{center}
\includegraphics*[width=8.5 cm,height=12.0 cm,angle=270]{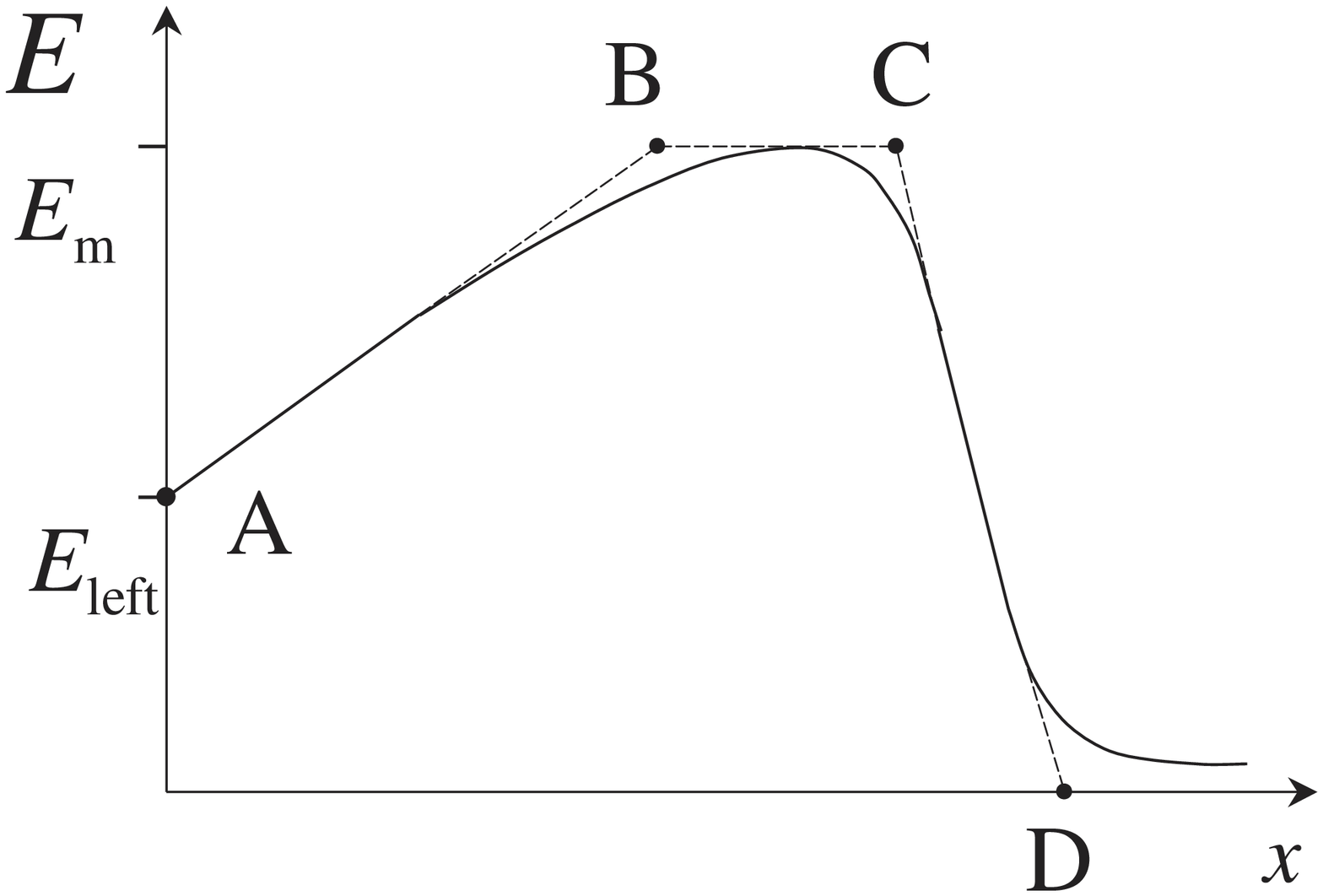}\\
\end{center}
\caption {Piece-wise linear approximation of the field
profile used to calculate the voltage $u$ across the $n$ base (Sec. 3H).
The respective  $\sigma(E)$ dependence
is shown by dashed line A--B--C--D in Fig.~\ref{sketch2}.}
\label{sketch3}
\end{figure*}

Basic parameters of plane impact ionization fronts in reversely
biased $p^{+}$-$n$-$n^{+}$ structure (Fig.\ \ref{sketch1}) are
determined by current density $J$ and concentration of initial
carriers $\sigma_0$ (regime parameters), doping of the $n$ base
$N_d$ (structure parameter) and such material parameters as
saturated drift velocities $v_{ns}$ and $v_{ps}$, low field
mobilities $\mu_n$ and $\mu_p$ and electron and hole impact
ionization coefficients $\alpha_n(E)$ and $\alpha_p(E)$. The front
velocity is given by $v_f \approx J/q N_d$ [Eq.\
(\ref{current3})]. The concentration of generated plasma
$\sigma_{\rm pl}$ and electric field in plasma $E_{\rm pl}$
determined by Eqs.\
(\ref{sigma_m},\ref{sigma_star},\ref{sigmaPlasma}) do not depend
on impact ionization coefficients $\alpha_{n,p}(E)$. Concentration
$\sigma_{\rm pl}$ weakly decreases with initial carrier
concentration $\sigma_0$ [Fig.\ \ref{FigSigmaPlasma}(a)]. For
moderate front velocities $v_f \lesssim 5 (v_{ns}+v_{ps})$ the
concentration $\sigma_{pl}$ and field $E_{\rm pl}$ are sensitive
to the ratio $v_{ns}/v_{ps}$, whereas the asymmetry in low-field
transport $\mu_{n}/\mu_{p} \ne 1$ has very little effect [Fig.\
\ref{FigSigmaPlasma}(b) and Fig.\ \ref{FigFieldPlasma}(b)]. For
higher front velocities $\sigma_{\rm pl}$ and $E_{\rm pl}$ do not
depend on $v_{ns}/v_{ps}$ and $\mu_n/\mu_p$ [Eq.\
(\ref{plasmaFast})]: $\sigma_{\rm}$ increases with $v_f$
quaislinearly whereas $E_{\rm pl}$ weakly decreases.

\begin{figure*}[hp]
\begin{center}
\includegraphics*[width=8.5 cm,height=12.0 cm,angle=270]{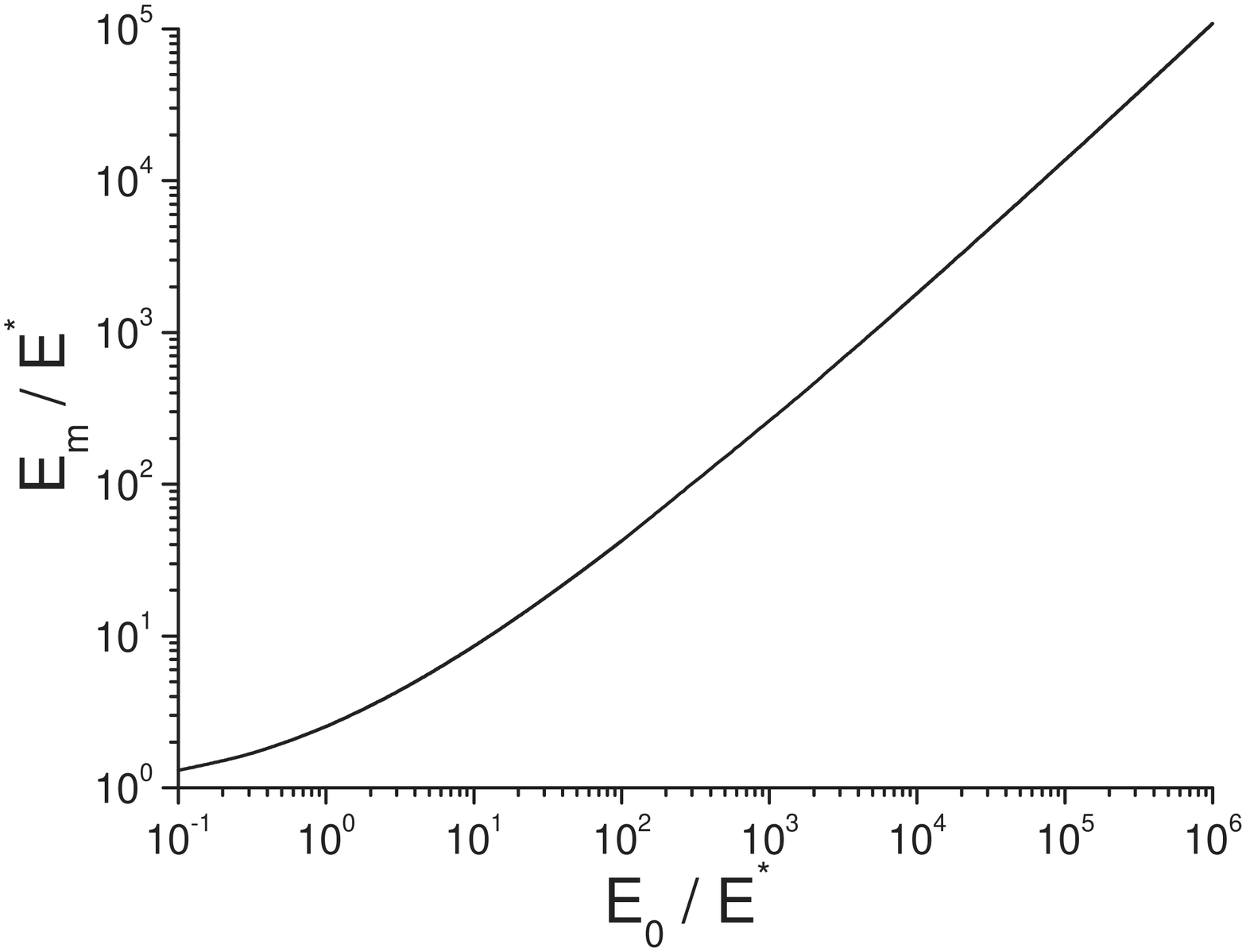}
\end{center}
\caption
{Maximum electric field $E_{\rm m}$
as a function of $E_0$  according to Eq.\ (\ref{E_m_Sym}).
Both $E_{\rm m}$ and $E_0$ are normalized by
$E^{\star}(v_f/v_s, \sigma_0/N_d, \alpha_0)$.
Note that $E_{\rm m}/E^{\star} \rightarrow 1$
when $E_0/E^{\star} \rightarrow 0$.}
\label{Figfield1}
\end{figure*}

\begin{figure*}[hp]
\begin{center}
\includegraphics*[width=8.5 cm,height=12.0 cm,angle=270]{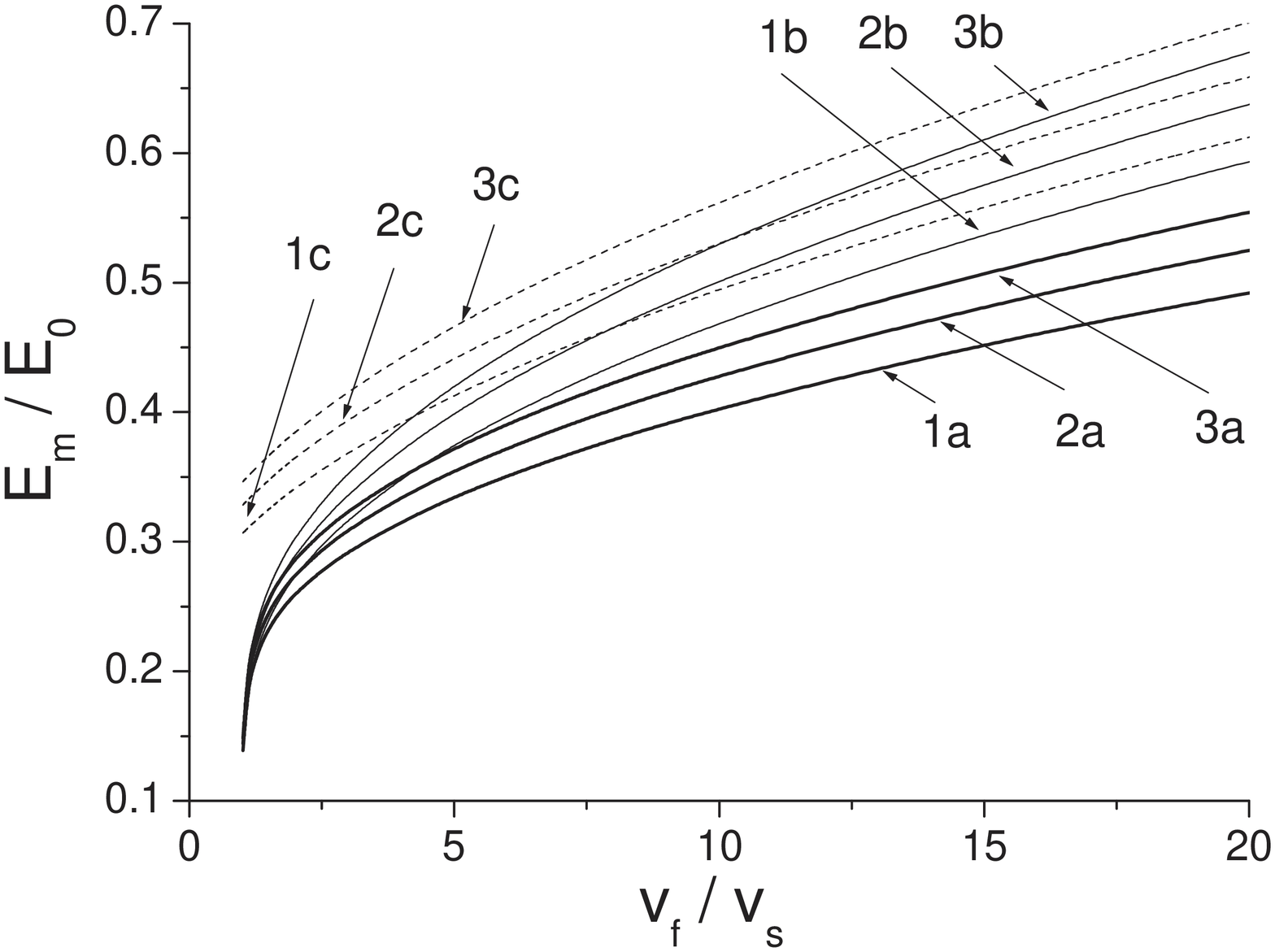}
\end{center}
\caption
{Maximum electric field $E_{\rm m}$ as a function of $v_f /v_s$ for different
values of $\sigma_0/N_d$ according to Eq.\ (\ref{E_m_Sym}).
Thick solid curves 1a,2a,3a correspond to the symmetric case (a)
$\alpha_n(E)=\alpha_p(E)$, $v_{ns}=v_{ps}$ [Eq.\ \ref{E_m_a}].
Thin solid curves 1b,2b,3b
correspond to impact ionization by electrons $\alpha_p(E)=0$,
$v_{ns}=v_{ps}$ [case (b), Eq.\ \ref{E_m_bc}],
dashed curves 1c,2c,3c correspond
to impact ionization by holes  $\alpha_n(E)=0$, $v_{ns}=v_{ps}$
[case (c), Eq.\ \ref{E_m_bc}].
Curves of 1st, 2nd and 3rd series correspond to
$\sigma_0/N_d=10^{-3},10^{-4}$ and  $10^{-5}$, respectively. The parameter
$b/(\alpha_0 E_0)=0.0002$
corresponds to the doping level $N_d=10^{14} \; {\rm cm^{-3}}$
in Si.}
\label{Figfield2}
\end{figure*}

\begin{figure*}[hp]
\begin{center}
\includegraphics*[width=8.5 cm,height=12.0 cm,angle=270]{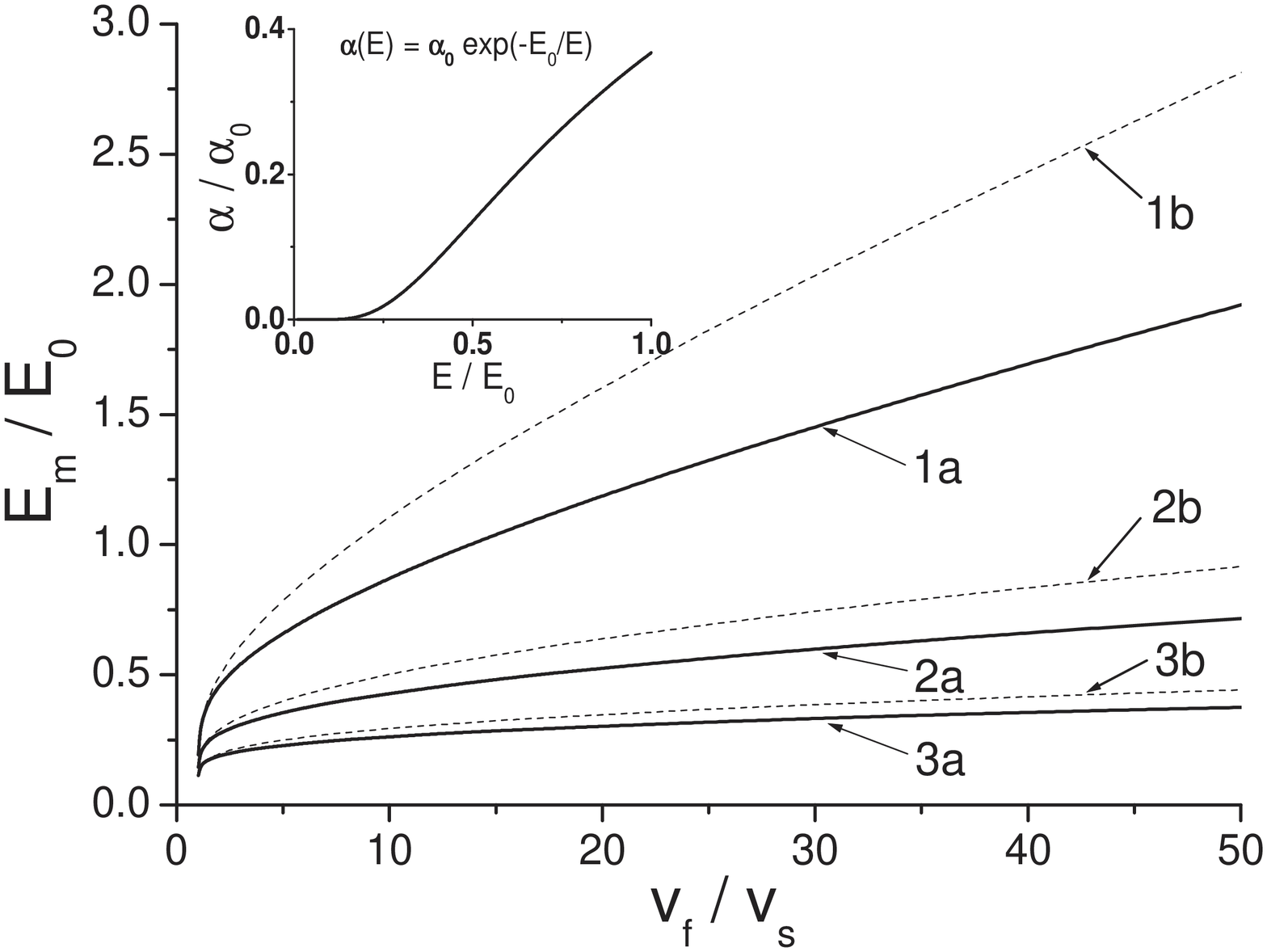}
\end{center}
\caption
{Maximum electric field $E_{\rm m}$ as a function of $v_f /v_s$ for different
values of $b/(E_0 \alpha_0)$ according to Eq.\ (\ref{E_m_Sym}).
Thick solid lines 1a,2a,3a correspond to the symmetric case
$\alpha_n(E)=\alpha_p(E)$, $v_{ns}=v_{ps}$ [case (a), Eq.\ (\ref{E_m_a})].
Thin solid lines 1b,2b,3b
correspond to impact ionization by electrons $\alpha_p(E)=0$, $v_{ns}=v_{ps}$
[case (b), Eq.\ (\ref{E_m_bc})].
Curves of 1st,2nd and 3rd series correspond to $b/(\alpha_0 E_0)=2\cdot 10^{-5},
2 \cdot 10^{-4}$ and $2 \cdot 10^{-3}$, respectively; $\sigma_0/N_d=10^{-4}$.
Insert shows the Townsends's dependence for impact ionization coefficient
$\alpha(E)$.}
\label{Figfield3}
\end{figure*}

\begin{figure*}[hp]
\begin{center}
\includegraphics*[width=8.5 cm,height=12.0 cm,angle=270]{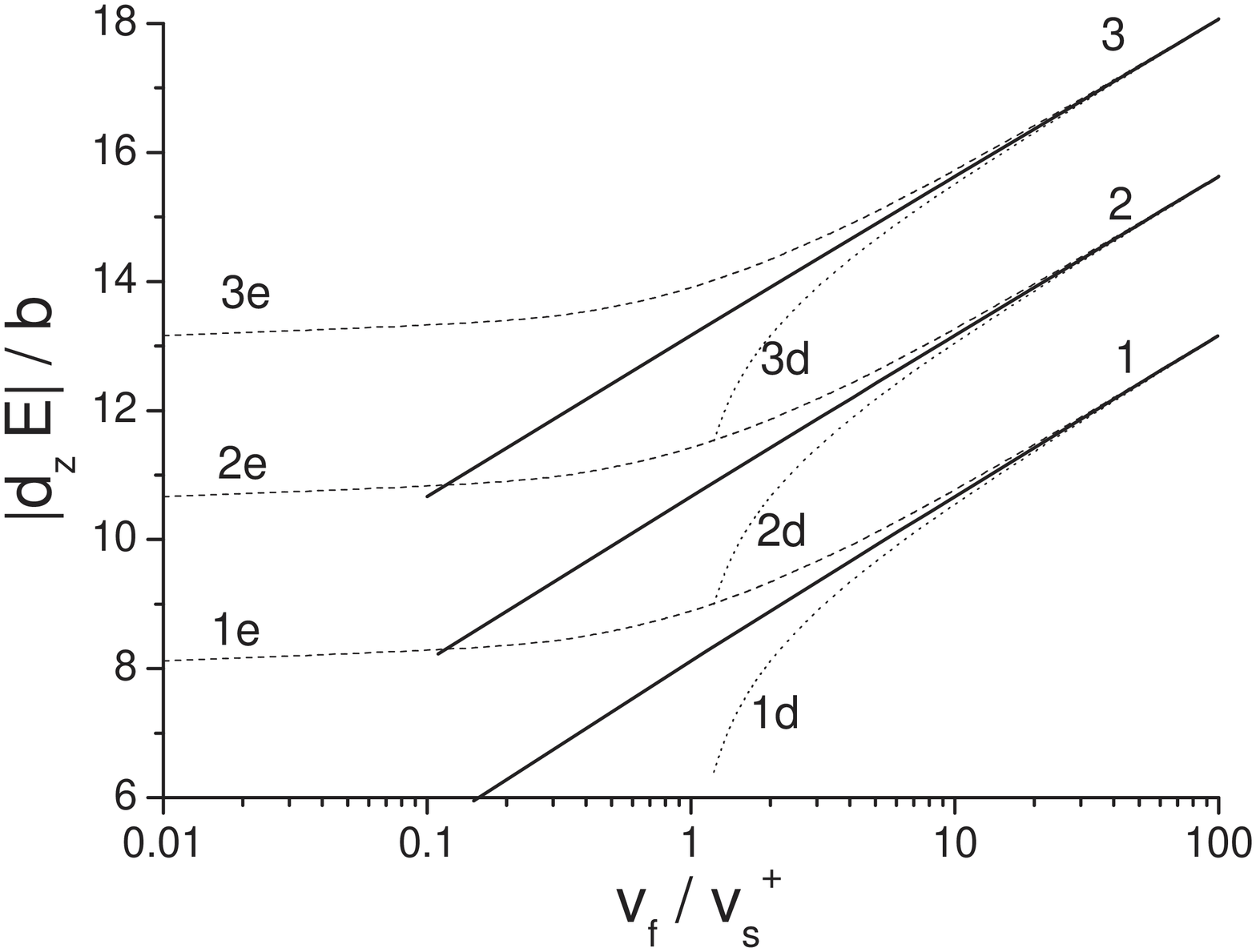}
\end{center}
\caption
{Slope of electric field in the screening region $|d_z E|$ normalized by the slope
in the depleted region $b=qN_d/\varepsilon \varepsilon_0$ as a function of $v_f/v_s^{+}$
for different values of $\sigma_0/N_d$.
Solid lines 1,2,3 correspond to the symmetric case $v^{-}_s=0$.
Dotted lines 1d,2d,3d and dashed lines 1e,2e,3e correspond to the two limiting
asymmetric cases  $v^{-}_s/v^{+}_s= - 1$ [immobile holes, case (d)] and
$v^{-}_s/v^{+}_s= +1$ [immobile electrons, case (e)], respectively.
Curves of 1st, 2nd and 3rd series correspond to
$\sigma_0/N_d=10^{-3},10^{-4},10^{-5}$, respectively.}
\label{fieldSlope}
\end{figure*}

General dependence of maximum electric field $E_{\rm m}$ on $v_f$
is given by Eq.\ (\ref{E_m_2}).
Due to strong nonlinearity of impact ionization coefficients $\alpha_{n,p}(E)$
often only one type of carriers contributes
to ionization. In this case the dependence $E_{\rm m}(v_f)$
can be determined in a simple form for the Townsend's approximation
$\alpha(E)=\alpha_0 \, \exp(-E_0/E)$
and symmetric transport $v_{ns}=v_{ps}$
[see Eq.\ (\ref{E_m_Sym}) and Fig.\ \ref{Figfield1}].
We reveal the existence of the effective threshold
of impact ionization $E_{\rm th} \approx 0.2 \, E_0$
(Figs.\ \ref{Figfield2} and \ref{Figfield3})
and the squareroot character of the $E_{\rm m}(v_f)$
dependence [Eq.\ (\ref{squareroot})]. Eq.\ (\ref{squareroot}) implies
that $v_f \sim {\ell_f}^2$, where $\ell_f$ is the effective width of
ionization zone. The squereroot dependence
fails for slow fronts when $\alpha_n \ll \alpha_p$.
$E_{\rm m}$ increases with $E_0$ and $N_d$ (Fig.\ \ref{Figfield3}) and decreases
with $\sigma_0$ and $\alpha_0$
(Figs. \ref{Figfield1},\ref{Figfield2},\ref{Figfield3}).
The width $\ell_{\rho}$ of the screening region where electric field falls from
$E_{\rm m}$ to $E_{\rm pl}$ weakly depends on $v_f$ and $\sigma_0$
[Eq.\ \ref{slope}]. The slope of the electric field in the screening
region is about 10...20 times larger than the slope
$q N_d/ \varepsilon \varepsilon_0$
in the depleted $n$ base the front propagates to (Fig.\ \ref{fieldSlope}).

The voltage over the structure $u$ is determined by the front velocity
$v_f$ and the front position $x_f$ (Eq.\ \ref{voltage3}).
The profile of electric field $E(z)$ is essentially triangular
since the nonlinear part near its maximum $E=E_{\rm m}$ is small.

In the case when the current density $J$ varies in time,
the front velocity $v_f$ and the front profile $E(z)$ are nonstationary.
The largest inner relaxation time $\tau_{\rho}=\ell_{\rho}/v_f$
is the time it takes for the front to travel over the width
of the screening region $\ell_{\rho}$ [Eq.\ (\ref{adiabatic1})].
The relations between basic front parameters
$E_{\rm}$, $\ell_{\rho}$, $\sigma_{\rm pl}$ and $E_{\rm pl}$
obtained for $J = {\rm const}$ remain valid if temporal variation of
$J$ is slow with respect to $\tau_{\rho}$ [Eq.\ (\ref{adiabatic2})].
For the actual case of the device connected in series
with an external load this adiabatic condition (\ref{adiabatic2}) can be presented
as $\ell_{\rho}/W \ll 1$ [Eq.\ (\ref{adiabaticLast})]:
inner and outer dynamics can be separated if ionization front is thin
with respect to the $n$ base width $W$. This condition is met or nearly met
for high-voltage sharpening diodes.

For very strong electric fields $E \gtrsim E_0$ the direct
band-to-band tunneling (Zener breakdown) must be
taken into account. Numerical simulations show that in presence of
this ionization mechanism the character of front propagation
substantially changes. \cite{ROD02a} The respective fronts have
been called tunneling-assisted impact ionization fronts.
\cite{ROD02a} Recently the dynamic avalanche breakdown of high voltage
diodes with stationary breakdown voltage $u_b \approx 1.5$~kV at
extremely high voltage about 10~kV that corresponds to electric
fields above the threshold of Zener breakdown has been observed
experimentally. \cite{RUK05} The analytical theory of
tunneling-assisted impact ionization fronts will be reported
separately.

\acknowledgements
We are grateful to P.~Ivanov for critical reading of the manuscript
and  helpful discussions. This work was supported by the Programm
of Russian Academy of Sciencies
``Power semiconductor electronics and pulse technologies".
P.R. thanks  A.~Alekseev for his hospitality at the University
of Geneva and acknowledges support from  the Swiss National
Science Foundation.



\begin{thebibliography}{99}
\bibitem{footnote1}
Generally, emission from deep-level centers can create nontrivial and nonstationary
profile of initial carriers in the $n$ base. However, our theory reveals
weak logarithmic dependence of all major front parameters on
$n_0$ and $p_0$. This justifies our assumption that $n_0$ and $p_0$ keep constant
values in the depleted region.
\end{thebibliography}
\end{document}